\documentclass[aps,prd,nofootinbib,superscriptaddress,preprintnumbers]{revtex4}

\pdfoutput=1
\usepackage{hyperref}
\usepackage{epsfig}
\usepackage{setspace}
\usepackage{textcomp}
\usepackage{amsmath}
\usepackage{graphicx}
\usepackage{slashed}
\usepackage{multirow}
\usepackage{amsmath,amsfonts,amssymb,graphicx,hyperref,hypcap}
\usepackage{subfigure}

\usepackage{color}

\usepackage{float}
\usepackage{tabu}

\newcommand{\ie}{{\it i.e.}}

\newcommand{\be}{\begin{equation}}
\newcommand{\ee}{\end{equation}}
\newcommand{\bea}{\begin{eqnarray}}
\newcommand{\eea}{\end{eqnarray}}
\newcommand{\ba}{\begin{array}}
\newcommand{\ea}{\end{array}}
\newcommand{\bi}{\begin{itemize}}
\newcommand{\ei}{\end{itemize}}
\newcommand{\bn}{\begin{enumerate}}
\newcommand{\en}{\end{enumerate}}
\newcommand{\bc}{\begin{center}}
\newcommand{\ec}{\end{center}}

\newcommand{\beq}{\begin{equation}}
\newcommand{\eeq}{\end{equation}}

\newcommand{\gsim}{\lower.7ex\hbox{$\;\stackrel{\textstyle>}{\sim}\;$}}
\newcommand{\lsim}{\lower.7ex\hbox{$\;\stackrel{\textstyle<}{\sim}\;$}}

\newcommand{\bs}{\begin{small}}
\newcommand{\es}{\end{small}}

\begin{document}

\title{Probing Dark-ALP Portals at Future $e^+e^-$ Colliders}
\vskip 1.0cm
\author{Sanjoy Biswas}
\email[]{sanjoy.biswas@rkmvu.ac.in}
\author{Anirban Chatterjee}
\email[]{anirban.chatterjee@rkmvu.ac.in}
\affiliation{Ramakrishna Mission Vivekananda Educational and Research Institute, 
Belur Math, Howrah 711202}
\author{Emidio Gabrielli}
\email[]{emidio.gabrielli@cern.ch}
\affiliation{Physics Department, Theoretical Section, University of Trieste, Strada Costiera 11, I-34151 Trieste,  and INFN, Sezione di Trieste, Via Valerio 2, I-34127 Trieste, Italy}
\affiliation{NICPB, Ravala 10, 10143 Tallinn, Estonia}
\affiliation{IFPU - Institute for Fundamental Physics of the Universe, Via Beirut 2, 34151 Trieste, Italy} 
\author{Barbara Mele}
\email[]{barbara.mele@roma1.infn.it}
\affiliation{INFN, Sezione di Roma,  P. le A. Moro 2, I-00185 Rome, Italy}

\vskip 1.0cm
\date{\today}
\vspace{0.5cm}
\begin{abstract}
\begin{center}
  {\bf Abstract}
\end{center}  
  We study portal interactions connecting  visible and  dark sectors, and involving 
  local interactions of a photon, a dark photon and a axion-like particle 
  (ALP)  at future $e^+e^-$ colliders. These interactions, mediated by higher-dimensional effective operators, may arise  
 at one-loop by  kinetic mixing between dark and ordinary photons, or, for massless dark photons, 
by direct short-distance contributions.  We explore these portal interactions  
for a heavy ALP  with masses between about 10 GeV and 230 GeV by investigating the sensitivity of the production
$e^+e^- \to \gamma  \gamma \bar{\gamma}$  to the effective couplings, where the dark photon $\bar\gamma$ gives rise to missing momentum in the final state. 
 We will show how an appropriate choice of missing-energy and missing-mass cuts can optimize the signal to standard-model background ratio. Exclusion regions for the effective  photon-dark-photon-ALP couplings versus the ALP mass are worked out  for a few representative  values of the collision energy and integrated luminosity, 
as presently envisaged by future $e^+e^-$ projects.
\end{abstract}
\maketitle

\section{Introduction} 
The persistent global consistency of the Standard Model (SM) predictions against the data collected so far at the CERN Large Hadron Collider (LHC) is radically changing our perspective on the origin of possible new physics beyond the SM, and on its characteristic energy scale.  This is also supported by the Higgs boson discovery \cite{Aad:2012tfa}, which is in good agreement with SM expectations \cite{ATLAS:2019slw,Sirunyan:2018koj}.
On the other hand, the growing evidences for dark matter (DM)~\cite{Ade:2013zuv}, and the lack of  suitable DM candidates in the SM framework are strengthening the possibility of the existence of  physics beyond the SM, provided DM does not have a purely  gravitational origin, as in the case of primordial black holes. A proper new physics model should be able to explain the correct dark-matter relic abundance and other experimental constraints coming from the direct and indirect searches for DM, while keeping itself well decoupled from ordinary matter in order to escape present TeV-range constraints in the LHC searches.

The present picture can motivate the idea that the new physics responsible for DM might reside in a hidden or Dark Sector (DS), the latter consisting of new particles which are singlets under the SM gauge interactions. 
The DS can eventually interact with the SM via some portal-type interactions, mediated by heavy messengers which can communicate tree-level interactions between the SM and the DS fields~\cite{Essig:2013lka}.
This mechanism can give rise to low-energy effective interactions  between SM and 
DS particles induced by higher-dimensional operators, the latter being suppressed by the characteristic scale of the messenger fields. Then, a quite heavy messenger sector might naturally explain why the new physics, and in particular the DM sector, is still escaping 
all direct and indirect searches~\cite{Essig:2013lka}.

A DS might contain a light or ultralight subsector. It might also be charged under its own long-distance interactions, in complete agreement with cosmological and astrophysical observations. Long-range forces mediated by a massless dark photon, corresponding to an exact $U(1)_D$ gauge symmetry in the DS, might also have a role in this picture. 
A DS scenario of the latter kind, which aims to solve the hierarchy puzzle of the SM Yukawa couplings, also providing natural DM candidates, has been recently proposed in \cite{Gabrielli:2013jka,Gabrielli:2016vbb}, implying a possible deep connection between the origin of Flavor and the DM interpretation.

Dark-photon  scenarios (both in the massive and massless cases) have been extensively considered in the literature in new-physics extensions of the SM gauge group \cite{Essig:2013lka},\cite{Holdom:1985ag}. They have  also been investigated in  cosmology and astrophysics~\cite{Spergel:1999mh,Vogelsberger:2012ku,Aarssen:2012fx,Tulin:2013teo,Ackerman:mha,Fan:2013tia,ArkaniHamed:2008qn,Zurek:2013wia}, mainly for improving models predictions.

In  collider physics, most of present dark-photon searches focus on  massive dark photons, where the broken $U(1)_D$ gauge field naturally develops  by  kinetic mixing a tree-level (milli-charged) interaction  with ordinary charged matter~\cite{Holdom:1985ag}. However, a massless dark photon can behave in a radically
different way with respect to a massive one. Indeed, the kinetic mixing among the ordinary photon and a massless dark photon can be rotated away, restricting  dark-photon interactions with ordinary matter to higher-dimensional operators~\cite{Holdom:1985ag}. Most of present  astrophysical and laboratory constraints applying to  massive dark photons can be evaded in the massless case, allowing for potentially large $U(1)_D$ couplings in the 
DS~\cite{Ackerman:mha}.

The DS could also contain the so-called Axion-Like Particles (ALP's), 
$a$, 
 loosely referring to  neutral light (or ultralight) scalar (or pseudo-scalar) particles. These particles can be present in 
SM extensions  motivated by the solution to the strong-CP problem (in which case the ALP is a QCD axion~\cite{Peccei:1977hh}) or can be  pseudo-Nambu-Goldstone bosons
corresponding to spontaneously broken continuous symmetries, either in the visible or in the DS, or a moduli field in string models \cite{Witten:1984dg,Svrcek:2006yi,Arvanitaki:2009fg,Acharya:2010zx}.

In the literature, the phenomenological aspects, including collider search of ALP's, have been extensively studied~\cite{Bauer1,Bauer2,Dolan,Brivio,Mimasu:2014nea,Jaeckel,Baldenegro:2018hng}.
 Most of these studies 
focus on the ALP effective coupling to  photons and/or gluons via the usual $a\, F^{\mu\nu} \!\tilde{F}_{\mu\nu}$ and $a \,G^{\mu\nu}\! \tilde{G}_{\mu\nu}$ types of interaction, 
involving the photon and gluon field-strength tensors, respectively.
In a possible theory UV completion, this kind of effective interactions could result (at one loop) from  integrating out some heavy messenger fields  connecting the dark and the observable sectors. Then the effective scale could be identified with the typical mass of the messengers running in the loop, properly rescaled by the product of internal couplings and  loop factor suppressions.

While fixed-target experiments~\cite{Jaeckel:2010ni,graham} and $B$  factories~\cite{Merlo:2019anv} are particularly useful for searching new weakly-coupled particles like ALP's in the MeV-GeV range,  high-energy colliders are more effective for constraining  ALP masses  above a few GeV's.  Collider investigations carried out so far in this context, involving the aforementioned dimension-five operators, mostly focused on {\it tri-photon} and/or 
{\it mono-photon+missing-energy} signatures (depending on the ALP stability  at the detector length scale) in the context of past and future $e^+e^-$ collider experiments as well as of (HL-)LHC experiments~\cite{Bauer1,Bauer2,Brivio,Mimasu:2014nea,Jaeckel,Baldenegro:2018hng}.

Motivated by the above scenarios, we analyze a different type of portal interactions, that is Dark-ALP portals which connect the visible sector and the DS via  a higher-dimensional effective operator $a F^{\mu\nu} \!\tilde{\bar{F}}_{\mu\nu}$, involving a photon $\gamma$, a dark photon  $\bar{\gamma}$
and an ALP $a$, being ${\bar{F}}_{\mu\nu}$ the dark-photon field strength.
The $a\gamma\bar\gamma$ interaction can arise from the
usual ALP coupling with photons, $a\, F^{\mu\nu} \!\tilde{F}_{\mu\nu}$, after rotating away the kinetic mixing in the photon dark-photon sector, or can directly be  induced at one loop by short-distance effects, after integrating out some heavy messenger fields. The same kind of interactions has been considered  in the context of various low-energy constraints in~\cite{kaneta1,kaneta,deNiverville:2018hrc}.

 We are now going to focus on a 
 massive ALP  scenario, with the $a$ mass $M_a$ in the range of $10\,{\rm GeV} \lsim M_a \lsim 230\,{\rm GeV}$, and a massless dark photon. 
Indeed, the new effective operator not only provides a rich phenomenology in the context of astrophysical and cosmological observations, including the possibility of low-energy observations, but 
also opens up new ALP search strategies  at collider experiments.
In particular, we propose to look at  the {\it di-photon plus missing-energy} channel $e^+e^- \to \gamma  \gamma \bar{\gamma} $
with the aim to probe both the $a\, F^{\mu\nu} \!\tilde{F}_{\mu\nu}$ and $a\, F^{\mu\nu} \!\tilde{\bar{F}}_{\mu\nu}$ types of effective operators, in the context of  future $e^+e^-$ collider experiments. 
As we will argue, this new channel not only  has a better  {\it signal-to-background} ratio  compared to the conventional {\it tri-photon} channel, even before imposing any hard 
cut on the final state objects. It also has more distinct kinematical features as compared to its SM background counterpart, thanks to the presence of a massless {\it invisible} dark-photon 
in the final state.
We will illustrate how such signal can be efficiently separated from the SM background by adopting various unconventional kinematic observables. 
In particular, we find that requiring an almost vanishing {\it missing mass} in the final state is particular effective in separating most of the SM background. 

While in the following we will assume new interactions involving a {\it  pseudoscalar} ALP coupling, our results can be easily generalized for a neutral {\it scalar} particle. 

The present search can be of relevance for the various $e^+e^-$ collider projects that are  presently under discussion,  in particular
for the linear colliders ILC~\cite{Bambade:2019fyw}, and CLIC~\cite{Charles:2018vfv}, and for the circular options
FCC-ee~\cite{Abada:2019zxq,Abada:2019lih} and CEPC~\cite{CEPCStudyGroup:2018ghi}. We will assume as reference for the
 collision center-of-mass (c.m.) energies, and integrated luminosities, the ones corresponding to the FCC-ee staging~\cite{Benedikt}. A straightforward projection
for the setup corresponding to a different machine can be done in most  cases.

We have organized the present paper in the following way. In the next section, we discuss the theoretical framework under consideration, and various constraints  relevant for the present study. In Section \ref{eventseln}, after presenting the LEP data implications on the present model, we suggest a new collider search strategy and  corresponding event selection criteria at future $e^+e^-$ colliders. Section \ref{results} contains the  results of our analyses. Finally, our conclusions are reported in Section \ref{conclusion}.

\section{Theoretical Framework}
A portal interaction connecting the dark sector and the visible sector can be parametrized in the following way
\bea
{\cal L}_{eff} = \frac{C_{a\gamma{\gamma}}}{\Lambda} a\, F^{\mu\nu} \!\tilde{F}_{\mu\nu} + 2 \frac{C_{a\gamma\bar{\gamma}}}{\Lambda} a\, F^{\mu\nu} \!\tilde{\bar{F}}_{\mu\nu},
\label{Leff}
\eea
where  $F^{\mu\nu}$ and $\bar{F}_{\mu\nu}$ are respectively the field strength of the photon $\gamma$ and the dark photon $\bar{\gamma}$, $a$ is the ALP field, 
${C_{a\gamma\bar{\gamma},a\gamma\bar{\gamma}}}$ are dimensionless couplings, and $\Lambda$ is some high energy scale\footnote{The factor 2 in  the second term has been introduced to take into account the effect of the Wick contractions in the first term arising from the matrix element with two external photon states.}~\cite{kaneta1}.
In principle, one could also add a term of the form $\frac{C_{a\bar{\gamma}\bar{\gamma}}}{\Lambda} a\, \bar{F}^{\mu\nu} \!\tilde{\bar{F}}_{\mu\nu}$ which might be present as well if one consider a specific UV completion of the low energy theory. In our  analyses this term doesn't play any role apart from modifying the ALP decay width, and hence the branching fractions of ALP decays.

In specific UV completions of the low energy theory, all these effective couplings can be related to each other. However, their connection will be in general  model dependent. 
For instance,
one could have an additional vertex involving the $Z$ boson, namely $\frac{C_{a \,Z\bar{\gamma}}}{\Lambda} a F_Z^{\mu\nu}\! \tilde{\bar{F}}_{\mu\nu}$, where $F_Z^{\mu\nu}$ is the  $Z$ field strength, related to the interactions in Eq.(\ref{Leff}). In the following, we  will take these couplings as independent parameters,  
assume that the contribution from  
$C_{a Z\bar{\gamma}}$  is negligible with respect to the two photon's operators, and consider only the effects of the two dominant terms 
in~Eq.(\ref{Leff}).
 
For our  collider analysis, we have considered ALP masses in the range $10 {\rm ~GeV} \lsim M_a \lsim \sqrt{s}$ , where $\sqrt{s}$ is the c.m. energy of $e^+e^-$ collisions. Actually, the existing experimental limits related to ALP searches are mostly sensitive 
to a light ALP [in particular, lighter than $\mathcal{O}$(GeV)], and come 
 from {\it e.g.} beam dump and light shining wall experiments, LSND and MinibooNE experiments, lepton $g-2$~\cite{Essig:2013lka}. 
Given the ALP mass range we  consider in  the present analysis, most of the stringent bounds on the axion coupling $C_{a\gamma\gamma}/\Lambda$ in Eq.(\ref{Leff}) carried out for very light  ALP's can be  neglected.
%

\section{Dark-ALP portal  with heavy ALP's at $e^+e^-$ colliders}
\label{eventseln}
In order to probe both the ALP-photon-photon,  $a\gamma{\gamma}$, and  
ALP-photon-dark-photon, $a\gamma\bar{\gamma}$, couplings, as defined in Eq.(\ref{Leff}), we study the
$e^+e^- \to \gamma\gamma\bar\gamma$ process at 
electron-positron colliders, where the dark photon $\bar\gamma$ behaves in the  detector just like a neutrino, {\it i.e.} giving rise to {\it massless}
missing momentum~\cite{Biswas:2015sha}.

 One of the main advantages of the {\it di-photon+missing-energy} channel over the conventional {\it tri-photon} channel (associated to the process $e^+e^- \to a\gamma\to \gamma\gamma\gamma$) is  the smaller SM background. 
As for illustration, at $\sqrt{s}\simeq M_Z$ the SM 
background in the {\it tri-photon} channel is about 4 pb with nominal cuts on the photon energy ($E_{\gamma}> 2$ GeV) and rapidity ($|\eta|<2.5$) along with the isolation between any 
photon pair ($\Delta R > 0.01$ where, $\Delta R = \sqrt{\Delta\eta^2 + \Delta\phi^2}$), whereas that in the {\it di-photon+missing-energy} channel coming from $\gamma\gamma\nu\bar{\nu}$ production is about 0.1 pb. 

As collision c.m. energy, we consider the  energies foreseen by the FCC-ee present staging, that is $\sqrt{s}\simeq [M_Z$,
160~GeV, 240 GeV], with integrated luminosity $L\simeq[150, 10, 5]$~ab$^{-1}$, respectively~\cite{Benedikt} (leaving aside here possible higher energy runs at 
$\sqrt{s}\simeq 365$ GeV). The corresponding results for the
ILC and  CEPC cases, and some extrapolation to the CLIC energies,  can be anyway obtained in a quite straightforward way from the present discussion.

\begin{figure}[H]
\begin{center}
\hskip -0.8cm \includegraphics[width=0.35\textwidth]{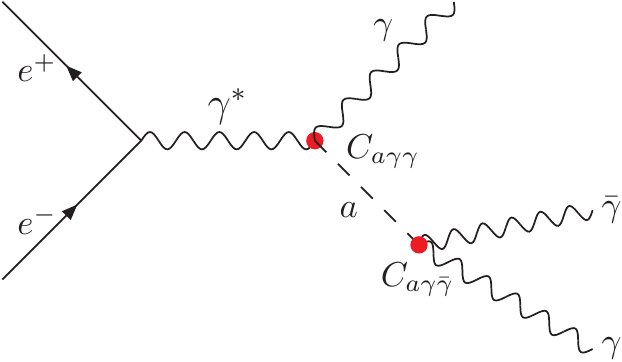} 
\hskip 1.2cm \includegraphics[width=0.35\textwidth]{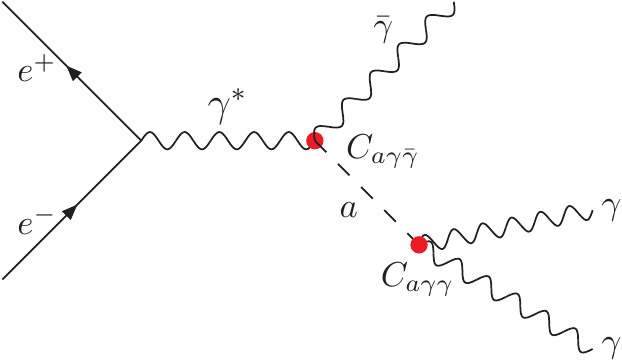} 
 \vskip 0.2cm
\caption{Feynman diagrams for the signal sub-processes $e^+e^- \to \gamma a \to \gamma \gamma \bar{\gamma}$ and 
$e^+e^- \to \bar{\gamma} a \to \gamma \gamma \bar{\gamma}$}
\label{FD}
\end{center}
\end{figure}

The $e^+e^- \to \gamma\gamma\bar\gamma$ process arises from the  couplings  in Eq.(\ref{Leff}) via the Feynman diagrams detailed in Figure \ref{FD}. The two diagrams 
interfere in a constructive way. Assuming $C_{a\gamma\gamma}=C_{a\gamma\bar{\gamma}}= 1$, the contribution due to interference varies from 4\% to about 17\% of the total cross section within the ALP mass range considered
in this analyses. The dominant contributions come from the incoherent sum of the individual sub-processes. 
The impact of interference effects can of course change as one moves away from the $C_{a\gamma\gamma}=C_{a\gamma\bar{\gamma}}= 1$ assumption   (a similar behaviour affects the ALP total width corresponding to the decays 
$a\to \gamma\bar{\gamma},\gamma{\gamma}$).

In the $e^+e^- \to \gamma\gamma\bar\gamma$ process, the ALP is produced on-shell, and it further undergoes a two body-decay to $\gamma\gamma$ or $\gamma\bar\gamma$. This 
 helps a lot  in the characterisation of the signal versus the SM background, which is mainly arising from the
 $e^+e^- \to \gamma\gamma\bar{\nu}\nu$ process, for an {\it invisible} dark photon. Indeed, 
one can treat the signal events as a $2\to 2\to 3$ process (modulo interference effects) and the momenta of the final state particles are  correspondingly constrained, with different constraints for different  subprocesses. On the other hand, the background $e^+e^- \to \gamma\gamma\bar{\nu}\nu$ is characterised by $2\to 4$ and  $2\to 3\to 4$ sub-processes, and the corresponding phase-space behaviour is in general significantly different from the signal one.  The presence of the missing momentum associated to a single invisible {\it massless} dark photon in the final state will provide a crucial additional handle to separate the signal 
from the SM background.

In the following we use the MadGraph5 event generator~\cite{Alwall:2014hca} to simulate both the signal and the background events.
We have implemented the effective ALP vertices using the FeynRules packages (v2.0)~\cite{Alloul:2013bka}. The output of the FeynRules (UFO model files) is  then interfaced with  
MadGraph5 (v2.6.3.2). 


\subsection{LEP Analysis}\label{seclep}

The effective $a\gamma\bar\gamma$ interactions between the ALP, photon and dark-photon can already be tested and constrained using the existing LEP data \cite{Gataullin:2005ge}. The L3
collaboration searched for single or multi-photon events with missing energy arising from $e^+e^- \to \bar{\nu}\nu \gamma (\gamma)$, in the c.m. energy range $\sqrt{s} = (189 - 208)$ GeV, 
with a total integrated luminosity of $619 ~{\rm pb}^{-1}$. In Table \ref{tablep}, we show the number of predicted signal events in our model, and the expected SM background 
events, along with the observed data in the {\it multiphoton+missing-energy} channel, after applying the set of cuts described in~\cite{Gataullin:2005ge}. 

\begin{table}[!h]
\large
\begin{center}
\tabulinesep=1.2mm
\begin{tabular}{|c|c|} 
\hline 
\hline
$M_a$ (GeV)  & $N_{\rm events} (n\gamma + \slashed{E})_{\rm LEP}$   \\
\hline
10 & 304 \\
\hline
50 & 810 \\
\hline
80 & 583 \\
\hline
150 & 77.8 \\
\hline
190 & 1.92 \\
\hline
\hline
expected (SM) & 115 \\
\hline
observed & 101 \\
\hline
\hline
\end{tabular}\\
\caption{Number of predicted signal events in our model (assuming $C_{a\gamma \gamma} = C_{a\gamma \bar{\gamma}} =1$  and   $\Lambda = 1$ TeV), in the {\it multiphoton+missing-energy} final state, with an integrated luminosity of 619 pb$^{-1}$, in the range $\sqrt{s} = (189-208)$ GeV, after applying the selection described in~\cite{Gataullin:2005ge}. 
The expected SM events and the observed ones are also presented in the last two rows, respectively, showing no excess in the data.}
\label{tablep} 
\end{center}
\end{table}

We have also estimated the corresponding $2\sigma$ exclusion limits on the model parameters (\ie, $C_{a\gamma \gamma} ~{\rm and}~ C_{a\gamma \bar{\gamma}}$, at fixed $M_a$ and $\Lambda$) coming from the LEP
analysis, and presented them in the following sections. Note that the LEP analysis of the 
{\it multiphoton+missing-energy} channel
has not been optimized for the $e^+e^- \to \gamma\gamma\bar\gamma$ process. In the next subsection, we will discuss the relevant  optimization strategy  in the context of
future $e^+e^-$ colliders. 

\subsection{Future $e^+e^-$ colliders}\label{secfccee}

\begin{figure}[H]
\begin{center}
\hskip -0.8cm \includegraphics[width=0.44\textwidth]{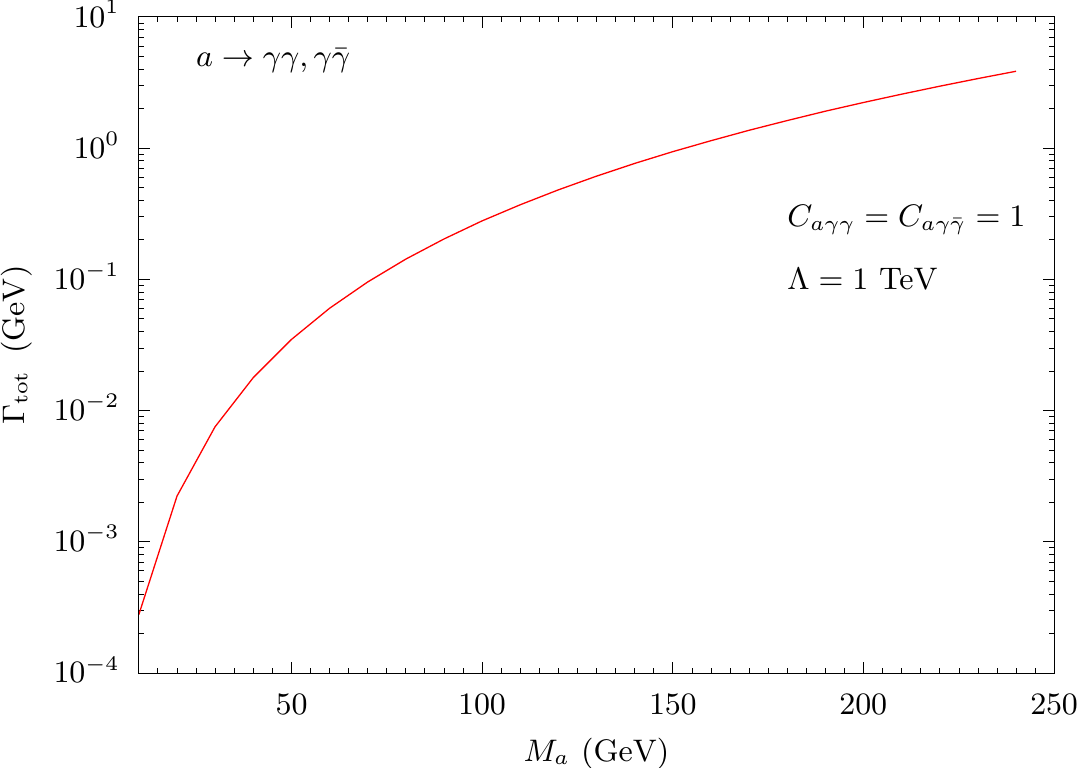} 
\hskip 1.2cm \includegraphics[width=0.44\textwidth]{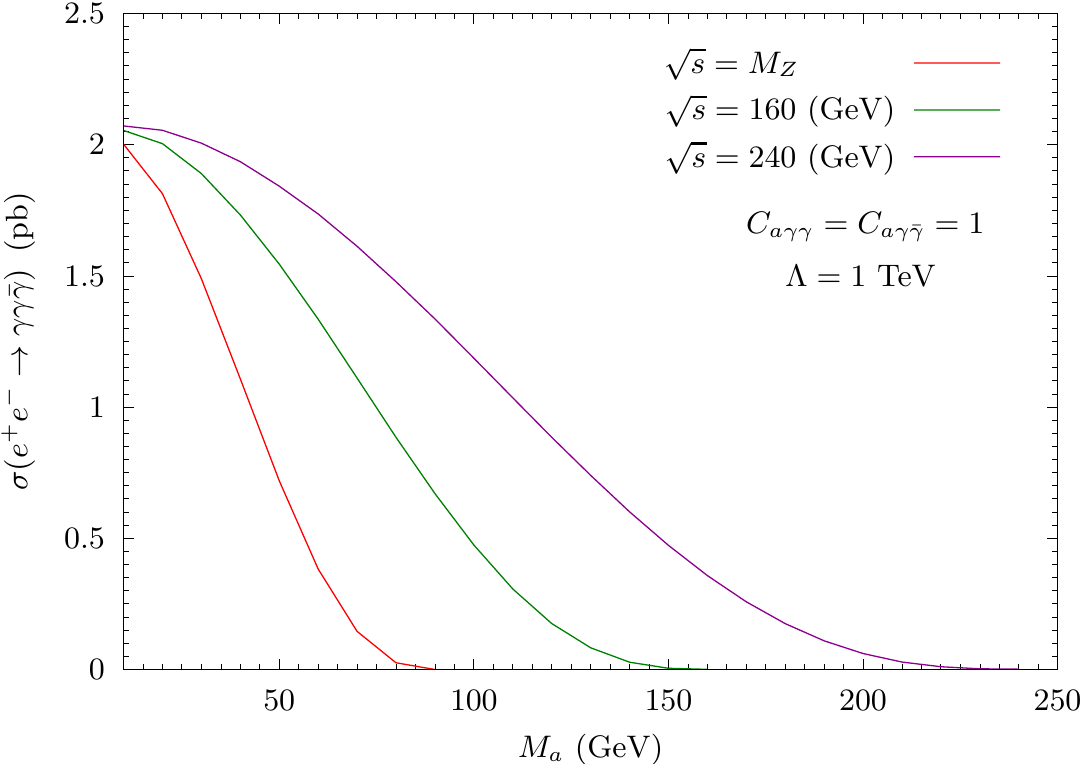} 
\caption{The left figure shows the variation of the ALP total decay width ($\Gamma_{\rm tot}$), as a function of the ALP mass $M_a$. The right  figure  shows  the total cross section  $\sigma(e^+ e^- \to \gamma \gamma \bar{\gamma})$  at different c.m. energies versus the ALP mass in  the  range $10 {\rm ~GeV} \lsim M_a \lsim \sqrt{s}$. We assume  $C_{a\gamma \gamma} = C_{a\gamma \bar{\gamma}} =1$  and   $\Lambda = 1$ TeV. The cross section presented 
 is the leading-order one,  not including the contribution from the process with extra photons  from initial-state radiation.}
\label{widthxsec}
\end{center}
\end{figure}

In this section we discuss the prospect of the $e^+ e^- \to \gamma \gamma \bar{\gamma}$ in the context of future $e^+e^-$ colliders. 
We start by discussing the signal cross section and the event selection criteria taking into account  various kinematical distributions.  
First we plot in Figure \ref{widthxsec} the ALP total decay width (left), and the total $e^+e^- \to \gamma\gamma\bar{\gamma}$ cross section (right) versus the ALP mass, assuming $C_{a\gamma\gamma} = C_{a\gamma\bar{\gamma}} = 1$ and $\Lambda = 1$ TeV. The left plot shows how, even for  ALP masses as low as 10 GeV,  the ALP is not stable on the detector 
length scale  down to couplings of the order $C_{a\gamma\gamma}, C_{a\gamma\bar{\gamma}} \sim 10^{-4}$,  
with $\Lambda\sim 1$ TeV, for which the decay length is of the order of 0.1 mm.
For narrower ALP widths, a displaced-vertices strategy might be in order.

\vskip 0.2cm

In order to account for  initial-state radiation effects, we have also added the radiative contribution of the $e^+e^- \to \gamma\gamma\gamma\bar{\gamma}$ and $e^+e^- \to \gamma\gamma\gamma\,\bar{\nu}\nu$ processes to the signal and background, respectively. 

The effect of finite detector resolution on the energy of each photon has been incorporated with a gaussian smearing function parametrised as in a typical ILC detector:
\be
\frac{\Delta E}{E} = \frac{16.6\%}{\sqrt{E/{\rm GeV}}} \oplus 1.1\%.
\ee

In order to reconstruct signal events as events containing at least 
 two isolated photons with some missing energy,  we impose the following set of 
 basic cuts on the final-state observables (in the following, we call  {C1} this  set of basic cuts for the  
{\it diphoton+missing-energy}  events): 
\bi
\item minimum energy for each photon, $E_{\gamma} > 2.$ GeV,
\item maximum rapidity for each photon, $|\eta_{\gamma}|<2.5$,
\item angular separation between any pair of photons  greater than $15^{\rm o}$,
\item minimum missing transverse energy, $\slashed{E}_T  > 5.$ GeV.
\ei

\begin{figure}[H]
\begin{center}
\hskip -2cm \includegraphics[width=0.49\textwidth]{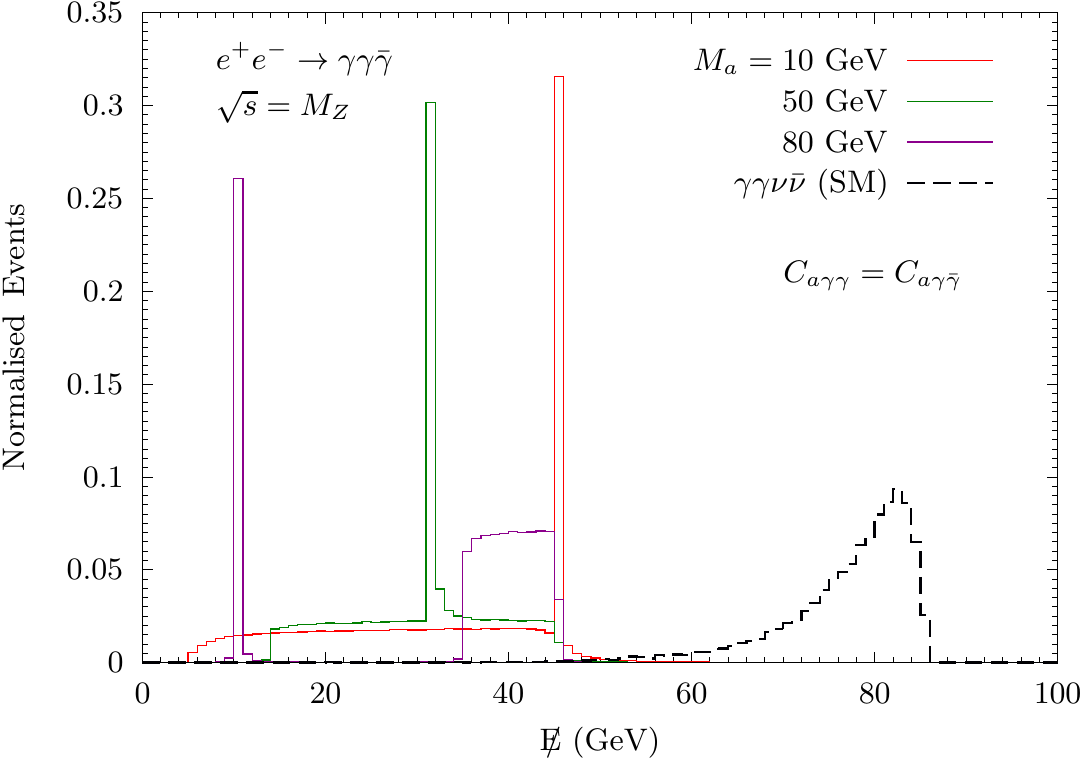} 
\hskip 0.5cm \includegraphics[width=0.49\textwidth]{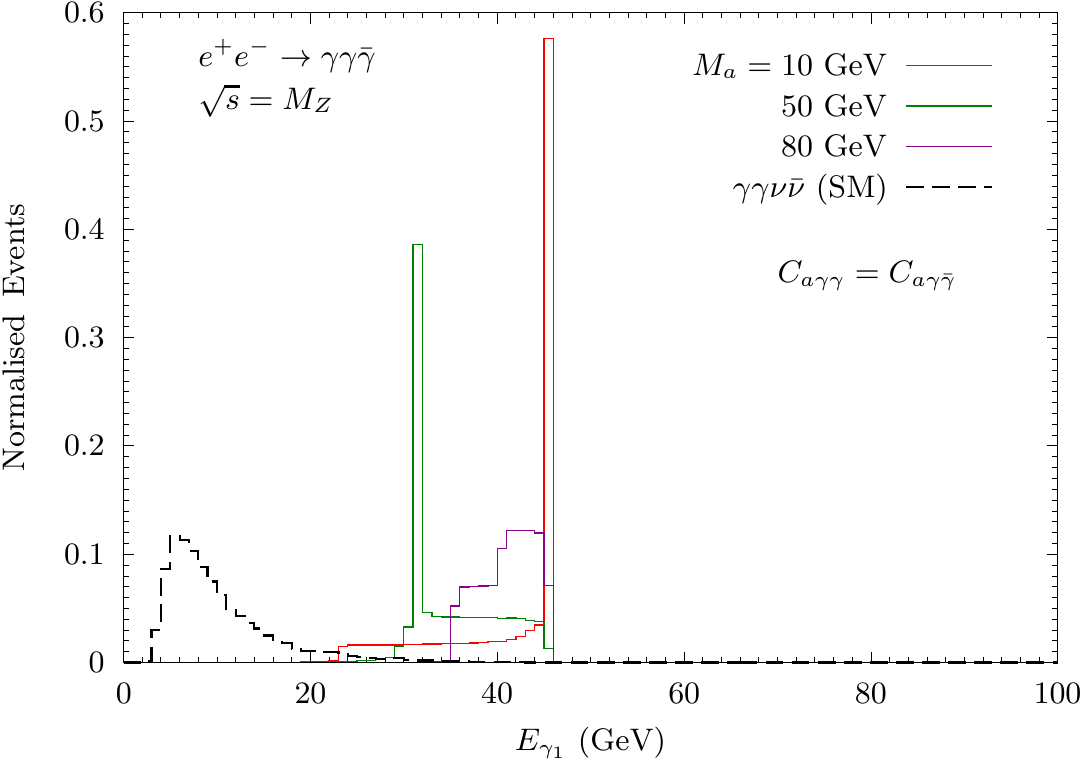} 
\caption{Missing energy ($\slashed{E}$) and  hardest-photon energy ($E_{\gamma_1}$) distributions
in the $e^+ e^- \to \gamma\gamma+\slashed{E}$ final state at  $\sqrt{s}\simeq M_Z$, 
for  a few  ALP-mass
benchmarks and the SM background ($\gamma\gamma\nu\bar{\nu}$). The basic cuts applied are the C1 set as defined in the text.}
\label{ecm91a}
\end{center}
\vskip 0.3cm
\end{figure}

\begin{figure}[H]
\begin{center}
\hskip -2cm \includegraphics[width=0.49\textwidth]{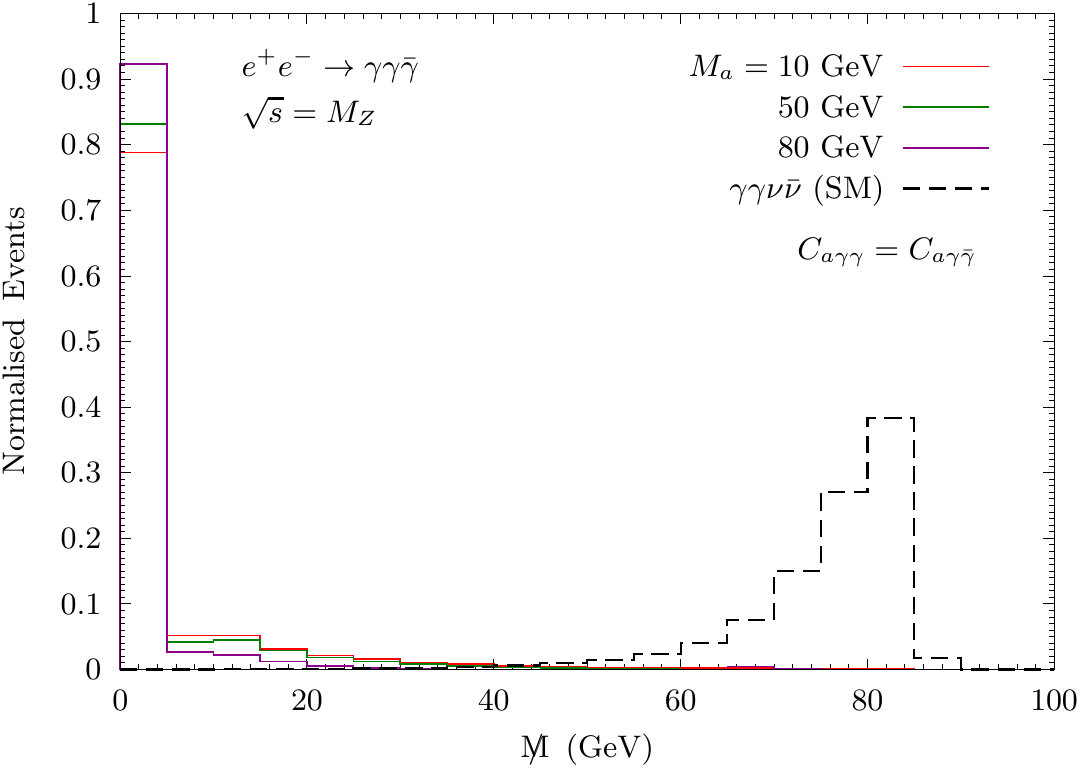} 
\hskip 0.5cm \includegraphics[width=0.49\textwidth]{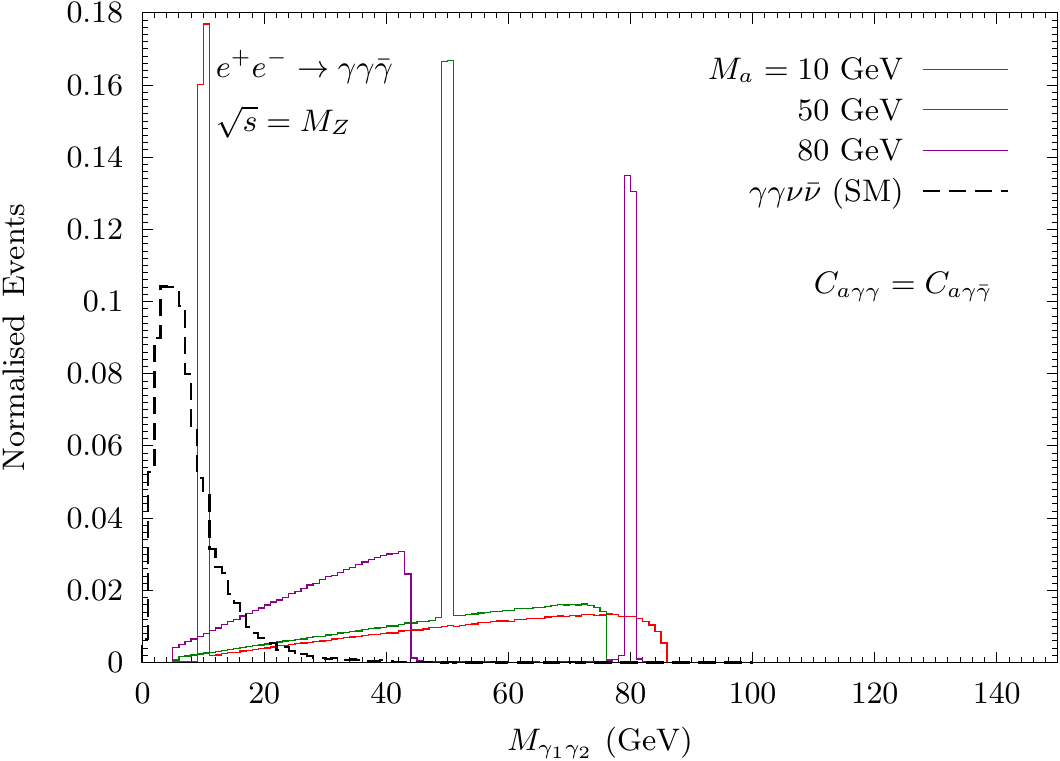} 
\caption{Missing mass ($\slashed{M}$) and diphoton invariant mass ($M_{\gamma\gamma}$) distributions
in the $e^+ e^- \to \gamma\gamma+\slashed{E}$ final state at $\sqrt{s}\simeq M_Z$, 
for  a few  ALP-mass
benchmarks and the SM background ($\gamma\gamma\nu\bar{\nu}$). The basic cuts applied are the C1 set as defined in the text.}
\label{ecm91b}
\end{center}
\end{figure}

\begin{figure}[H]
\begin{center}
\hskip -2cm \includegraphics[width=0.49\textwidth]{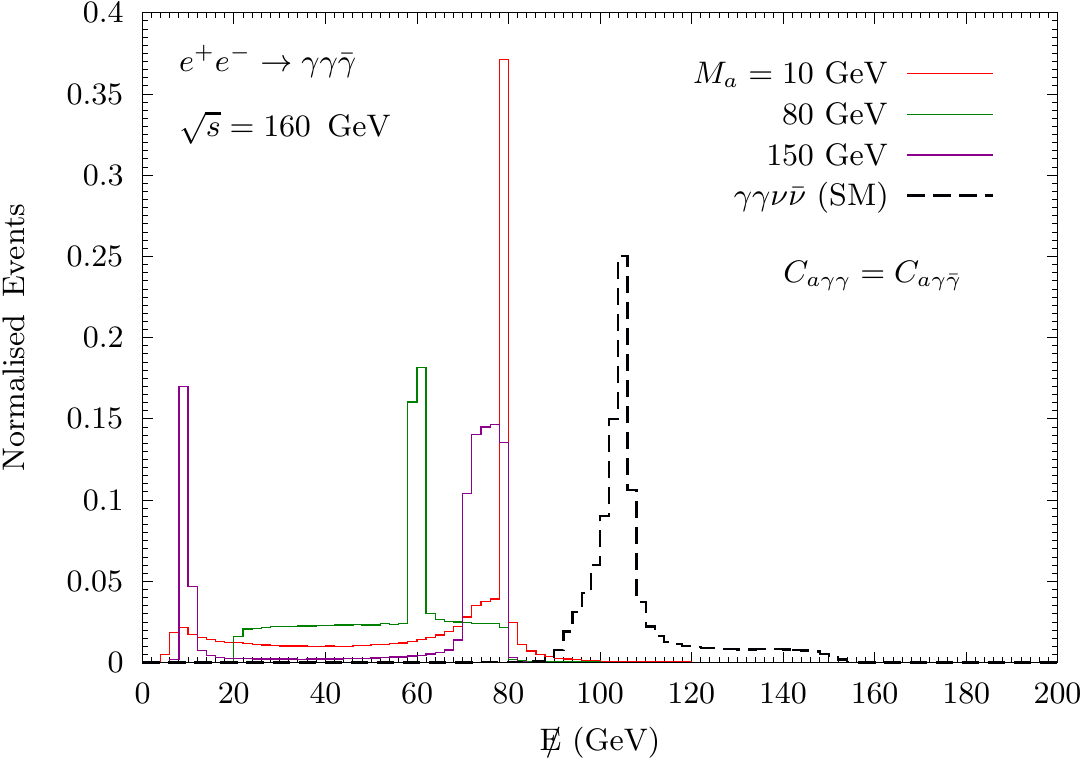} 
\hskip 0.5cm \includegraphics[width=0.49\textwidth]{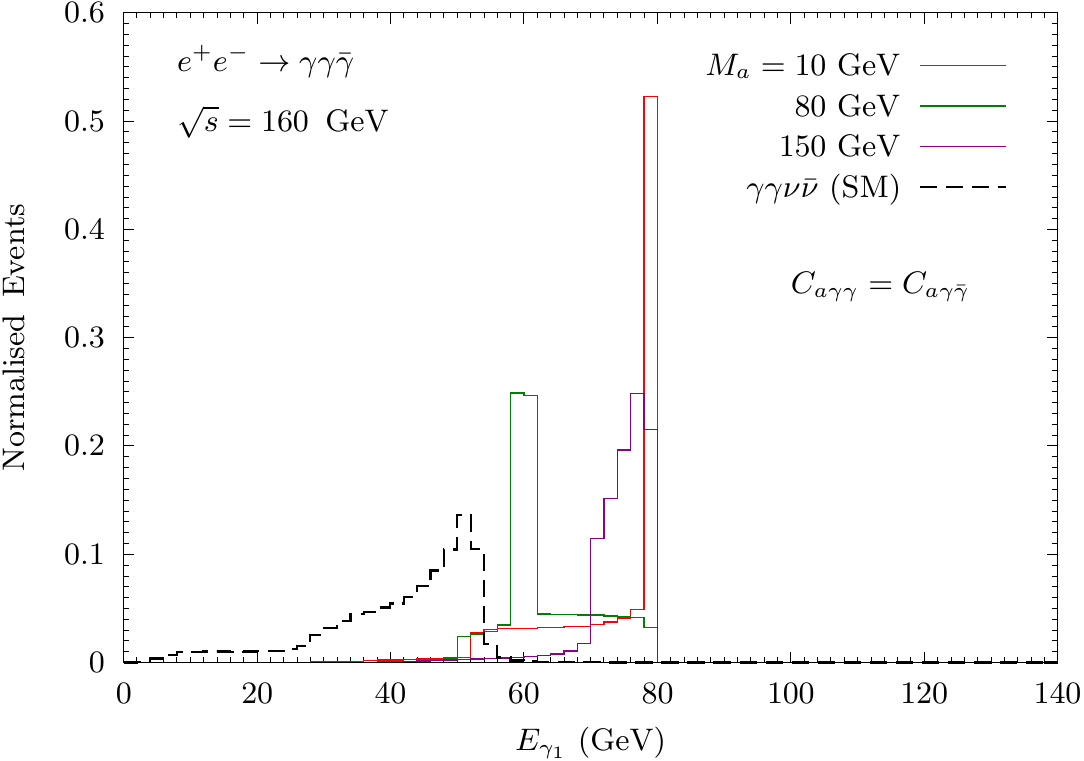} 
\caption{Missing energy ($\slashed{E}$) and  hardest-photon energy ($E_{\gamma_1}$) distributions
in the $e^+ e^- \to \gamma\gamma+\slashed{E}$ final state at  $\sqrt{s}=160$ GeV, 
for  a few  ALP-mass
benchmarks and the SM background ($\gamma\gamma\nu\bar{\nu}$). The basic cuts applied are the C1 set as defined in the text.}
\label{ecm160a}
\end{center}
\end{figure}

\begin{figure}[H]
\begin{center}
\hskip -2cm \includegraphics[width=0.49\textwidth]{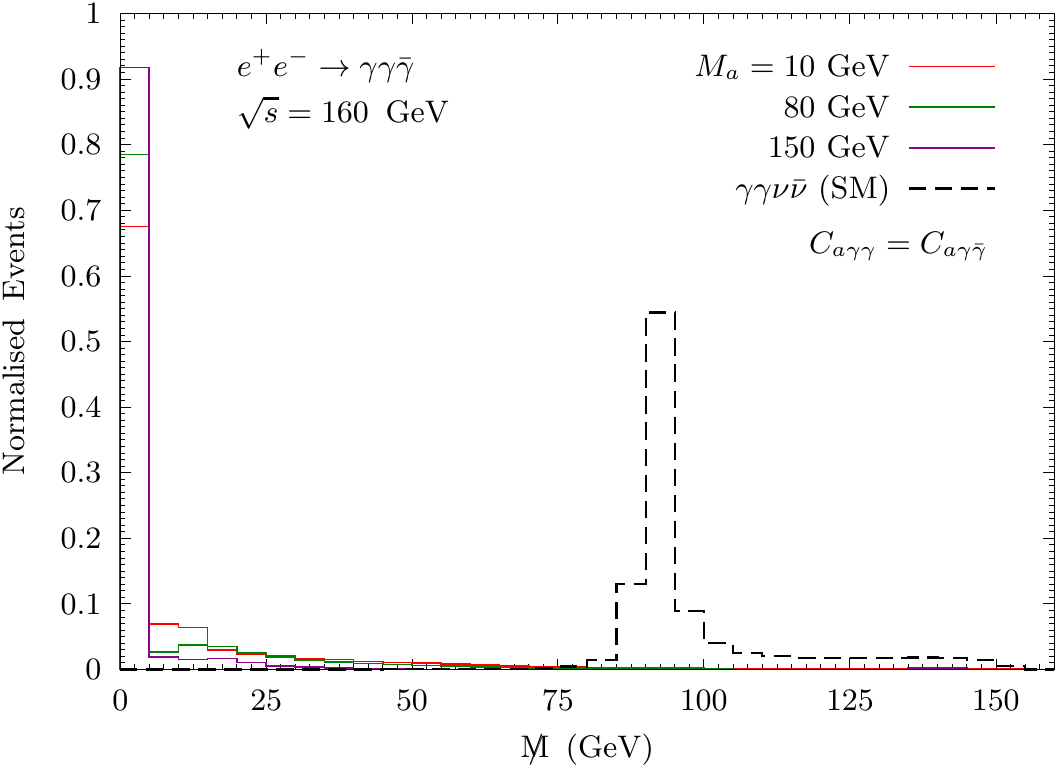} 
\hskip 0.5cm \includegraphics[width=0.49\textwidth]{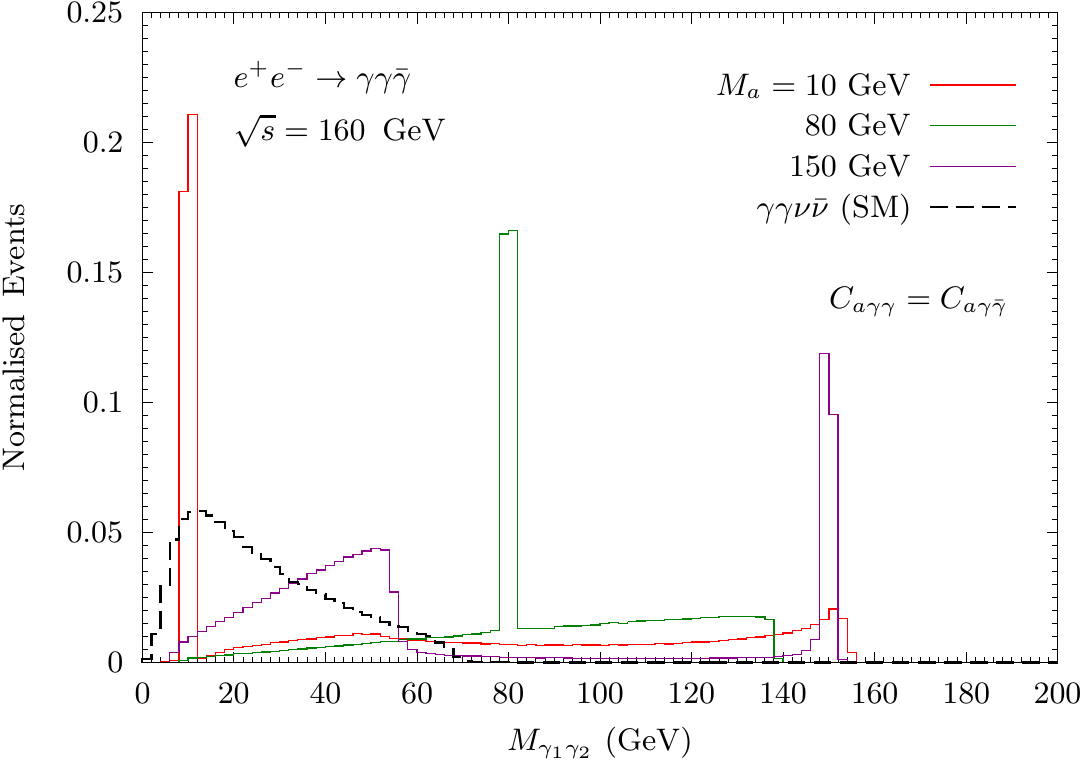} 
\caption{Missing mass ($\slashed{M}$) and diphoton invariant mass ($M_{\gamma\gamma}$) distributions
in the $e^+ e^- \to \gamma\gamma+\slashed{E}$ final state at  $\sqrt{s}=160$~GeV, 
for  a few  ALP-mass
benchmarks and the SM background ($\gamma\gamma\nu\bar{\nu}$). The basic cuts applied are the C1 set as defined in the text.}
\label{ecm160b}
\end{center}
\end{figure}

\begin{figure}[H]
\begin{center}
\hskip -2cm \includegraphics[width=0.49\textwidth]{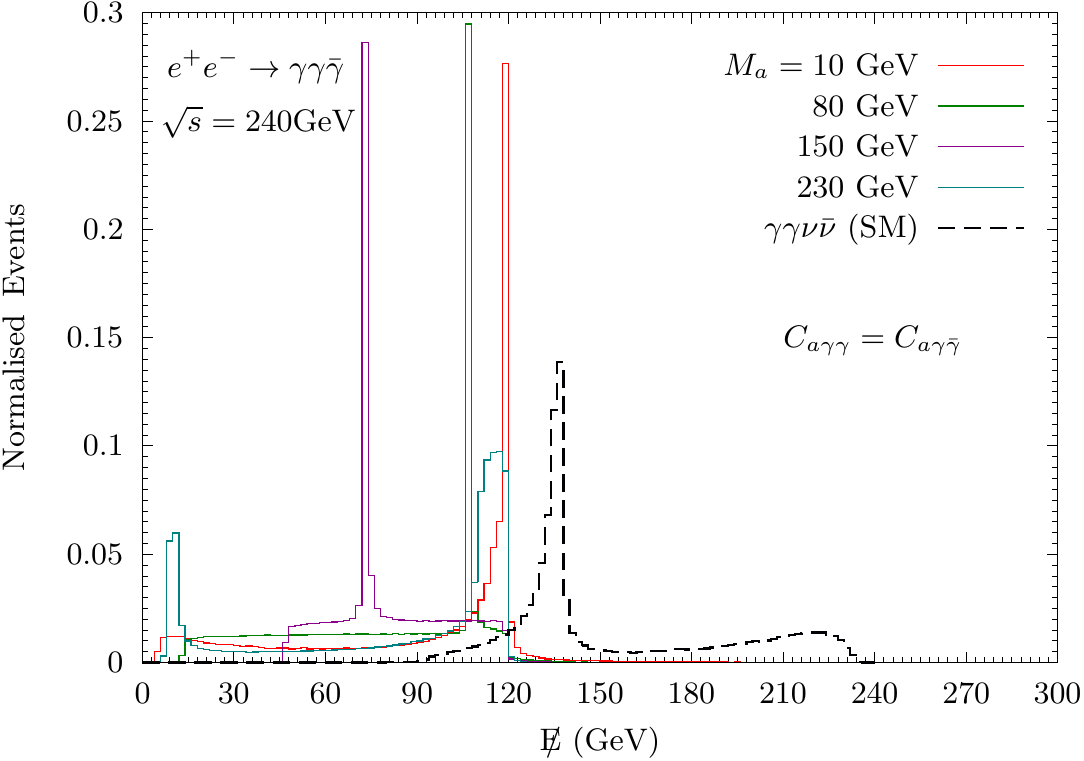} 
\hskip 0.5cm \includegraphics[width=0.49\textwidth]{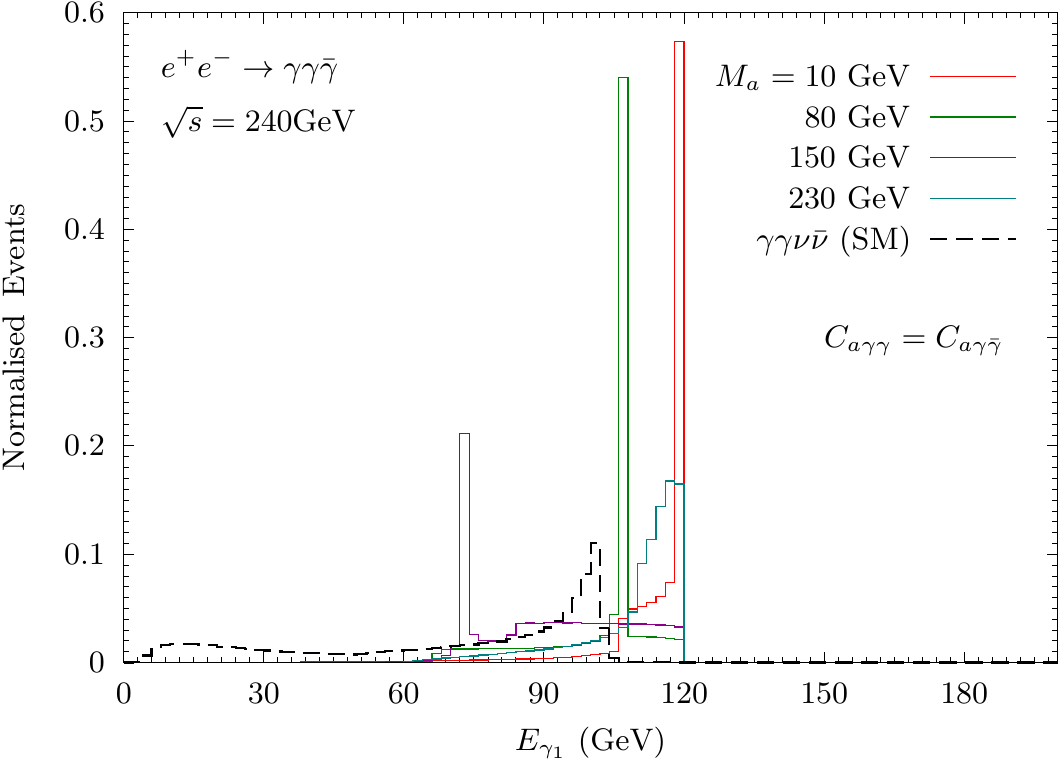} 
\caption{Missing energy ($\slashed{E}$) and  hardest-photon energy ($E_{\gamma_1}$) distributions
in the $e^+ e^- \to \gamma\gamma+\slashed{E}$ final state at  $\sqrt{s}=240$ GeV, 
for  a few  ALP-mass
benchmarks and the SM background ($\gamma\gamma\nu\bar{\nu}$). The basic cuts applied are the C1 set as defined in the text.}
\label{ecm240a}
\end{center}
\end{figure}

\begin{figure}[H]
\begin{center}
\hskip -2cm \includegraphics[width=0.49\textwidth]{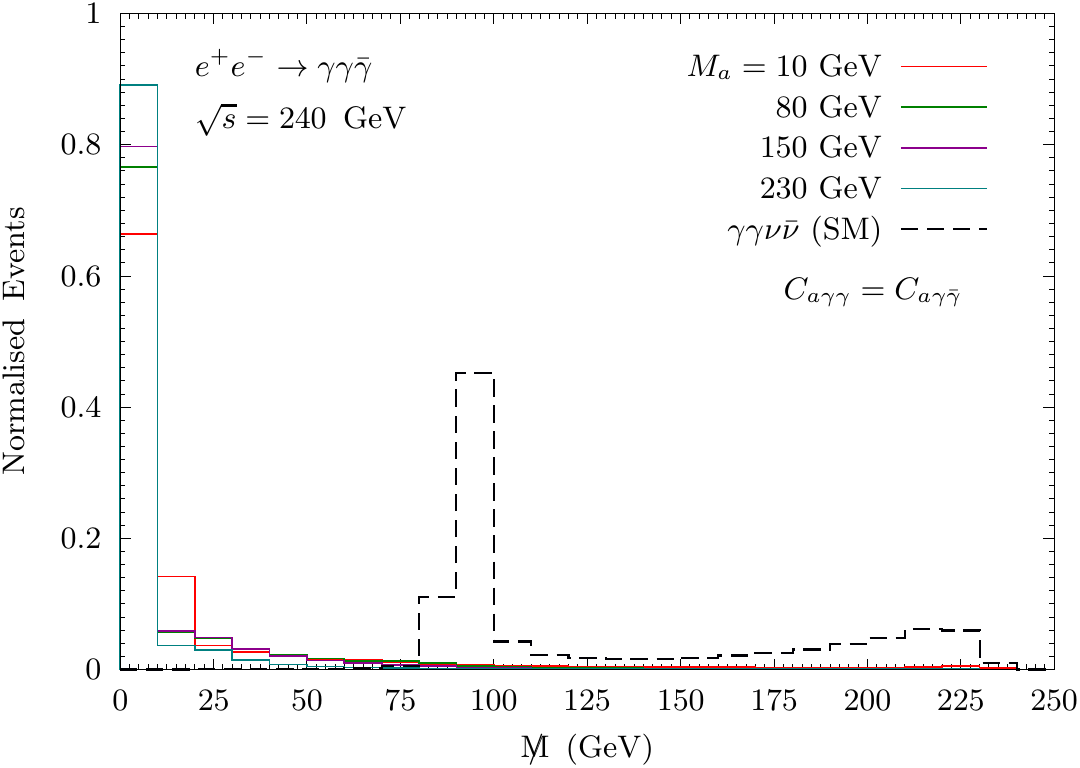} 
\hskip 0.5cm \includegraphics[width=0.49\textwidth]{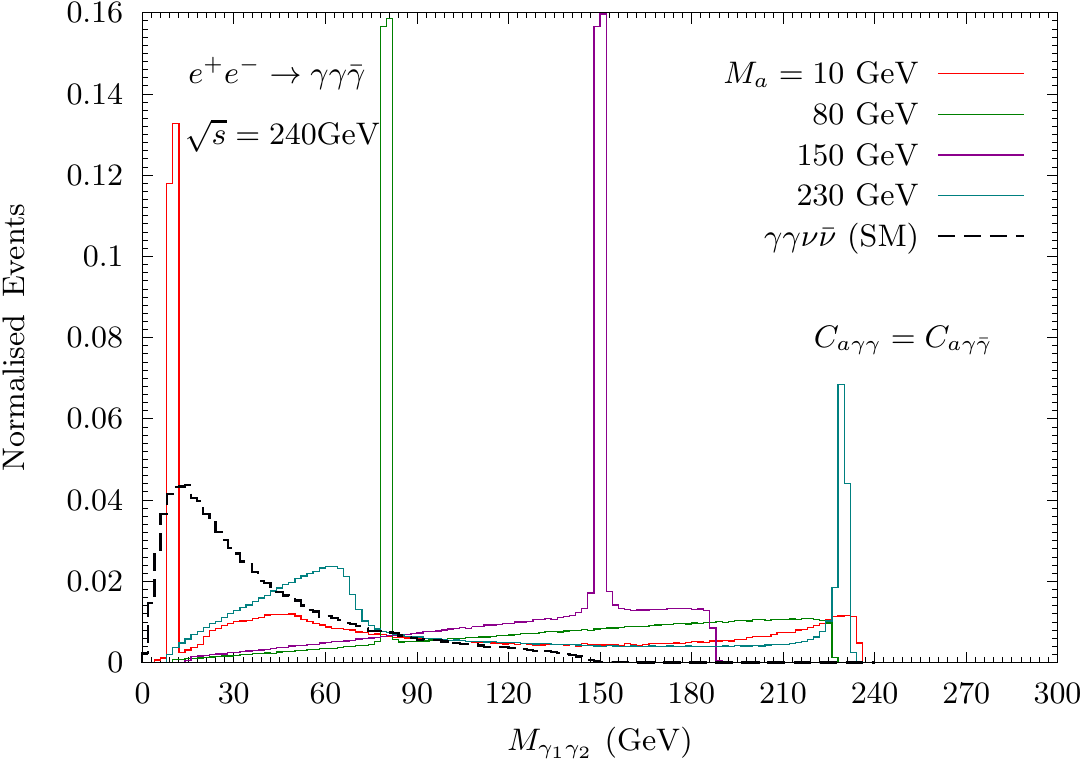} 
\caption{Missing mass ($\slashed{M}$) and diphoton invariant mass ($M_{\gamma\gamma}$) distributions
in the $e^+ e^- \to \gamma\gamma+\slashed{E}$ final state at  $\sqrt{s}=240$ GeV, 
for  a few  ALP-mass
benchmarks and the SM background ($\gamma\gamma\nu\bar{\nu}$). The basic cuts applied are the C1 set as defined in the text.}
\label{ecm240b}
\end{center}
\end{figure}

In Figure \ref{ecm91a}, we plot the normalised distributions for the missing energy $\slashed{E}$ associated to the dark photon (or, in the background case, to the $\bar{\nu}\nu$ system), and for the 
hardest-photon energy   $E_{{\gamma}_1}$, for  signal and  background, at a c.m. energy $\sqrt{s}\simeq M_{Z}$. Figure \ref{ecm91b} shows the missing mass $\slashed{M}$ distribution associated to the invisible system (with  
$\slashed{M}$ defined below), 
and the diphoton-system invariant mass distribution, for  signal and background. The same is shown in Figures \ref{ecm160a},  \ref{ecm160b}, \ref{ecm240a} and \ref{ecm240b} at different  c.m.  energies, namely, $\sqrt{s}=160$ GeV and 240 GeV. All these 
 distributions have been obtained after applying the set of basic cuts {C1}, mentioned above. 
 
 In general, the shape of the $\slashed{E}$ distributions for the signal  contains a peak superimposed on a box distribution. This  corresponds to the different kinematics associated the two subprocesses where either a $\gamma\bar\gamma$ or a $\gamma \gamma$ system is resonating at the ALP mass (see Figure \ref{FD}). 
In one case, the ALP  is produced in association with a photon, giving rise to the box distribution for the dark photon which arises from the ALP decay. In the second case,
the ALP  is produced in association with the dark photon, which is essentially monochromatic. The peak position corresponding to the monochromatic component in the $\slashed{E}$ distribution is given by
\be
\slashed{E}_{peak} \simeq \frac{\sqrt{s}}{2}(1-\frac{M_a^2}{s}),
\ee
while the minima and maxima of the box distributions are given by 
\be
\slashed{E}_{min}  \simeq \frac{1}{2}\frac{M_a^2}{\sqrt{s}} \; , \;\;\; 
\slashed{E}_{max}  \simeq \frac{\sqrt{s}}{2}.
\ee

When $M_a\approx \sqrt{s}$, the distribution  splits up into two separate kinematical regions for a given $M_a$, 
because $\slashed{E}_{peak}\ll \slashed{E}_{min}\approx \slashed{E}_{max}$.  On the other hand, for $0 \lsim M_a \lsim \sqrt{s/2}$, the peak lies between the 
$\slashed{E}_{min}$ and $\slashed{E}_{max}$ ($\slashed{E}_{min} < \slashed{E}_{peak} < \slashed{E}_{max}$). 
This is clearly shown in Figures \ref{ecm91a}, \ref{ecm160a}, and \ref{ecm240a}.

 \vskip 0.3cm
\vskip -0.5cm

\vskip -0.5cm

In addition to the $\slashed{E}$ variable, we have introduced   the invariant mass of the invisible system~\cite{Biswas:2015sha}. The invariant mass of the invisible system 
(or {\it missing mass}) $\slashed{M}$  is defined as
\be
\slashed{M} = \sqrt{\slashed{E}^2 - \slashed{\vec{P}}^2} ,
\ee
and turns out to be a crucial variable to separate the signal from the SM background. 

Indeed, the signal  invisible momentum is carried by the dark photon which is massless in our  scenario. Therefore the missing mass is 
expected to peak near zero. On the contrary, the background  invisible momentum is carried by the $\bar{\nu}\nu$-system for which the mass  distribution is
expected to be dominant at quite large values. 

We extensively make use of the  $\slashed{E}$ and $\slashed{M}$  kinematical variables to separate the signal from the SM background. To this end, we set an upper limit on the missing energy and the missing mass associated with the accepted events.  These two variables are quite independent, unless one fixes the invariant mass of the diphoton system $M_{\gamma\gamma}$, 
on which we actually do not put any constraint.  

On the basis of the above distributions, we will impose 
 a cut $\slashed{M}< 10$ GeV, and a cut  $\slashed{E} < \sqrt s/2 $,
in order to optimize the signal significance, for the following choise   of  $\sqrt{s}$ and  $M_{a}$ parameters:

%
\vskip -0.5cm

\bi
\item $\sqrt{s}\simeq M_{Z}$, $10 {\rm ~GeV} \lsim M_a \lsim 80 {\rm ~GeV}$; 


\item $\sqrt{s}=160$  GeV, $10 {\rm ~GeV} \lsim M_a \lsim 150 {\rm ~GeV}$; 


\item $\sqrt{s}=240$  GeV, $10 {\rm ~GeV} \lsim M_a \lsim 230 {\rm ~GeV}$.  


\ei




\vskip 0.5cm
\section{Results and discussion}
\label{results}

We  now present the results of our analyses. The final choice of the  event selection criteria is a natural consequence of the
kinematic distributions discussed in the previous section. The combination of the missing energy $\slashed{E}$ and missing 
mass $\slashed{M}$ cuts reduces the SM background substantially. For $\sqrt{s}\simeq M_{Z} $ and 160~GeV, the optimized
choice of cuts is independent of the ALP masses. This is because both the missing-energy and the missing-mass distributions have very 
small or almost negligible overlap with background events. 
On the other hand, on the basis of $\slashed{E}$ distributions, 
 a mild $M_a$ dependence is expected  at $\sqrt{s}=240$~GeV. However,
 we could obtain a significance very close to one corresponding  to the $M_a$ dependent optimized cuts, by adopting the same (rescaled) cut choice as used at lower $\sqrt s$.

In Tables \ref{eventtab1}, \ref{eventtab2} and \ref{eventtab3}, we present the numerical results of our analysis for a few representative $M_a$ benchmarks in order to illustrate the 
effects of the cut-flow  on the signal and the background events discussed  earlier, at various $e^+e^-$ collision energies. The signal events reported correspond to  $C_{a\gamma\gamma} = C_{a\gamma\bar{\gamma}} = 1$ and $\Lambda = 1$ TeV, and can be easily rescaled for different couplings and energy scale. We name $\sigma_{\rm cut}$, the residual cross sections
after applying the complete cut-flow to the process phase-space.

\vskip 0.4cm

\begin{table}[!h]
\large
\begin{center}
\tabulinesep=1.2mm
\begin{tabular}{|l|c|c|c|c|c|} 
\hline 
\hline
\multicolumn{2}{|c|}{$\sqrt{s}\simeq M_Z$} & 
\multicolumn{3}{|c}{~~~~~~~~~~~~$N_{\rm events}$}& \\
\hline
$M_a$ (GeV)  & $\sigma_{\rm tot}$ (fb) &      Basic cuts & $\slashed{E}< 46$ GeV &  $\slashed{M}< 10$ GeV & $\sigma_{\rm cut}$ (fb)   \\
\hline
~ ~10  &  ~2002~    & 839514   & 813996 &  704647 & 1410    \\ 
\hline 
~ ~50  &  ~719~   & 935228   &  929024   &  813901 & 585  \\  
\hline
 ~ ~80  &  ~25.9~    &  877285   &  874469   & 833578 & 21.7 \\ 
\hline
\hline
 ~  $\gamma\gamma \nu \bar{\nu}$    & ~2544~  &    12884  &  44  & 1 & 0.0025   \\ 
\hline 
\hline
\end{tabular}\\
\caption{Cross sections (before and after cuts) and event counts out of $10^6$ simulated events, for the signal   and 
the SM $\gamma\gamma \nu \bar{\nu}$ background at  $\sqrt s\simeq M_{Z}$ (versus $M_a$), applying the C1 set of basic cuts as described in the text, followed by further sequential cut optimization. We assume  $C_{a\gamma\gamma}=C_{a\gamma\bar{\gamma}}=1$
and $\Lambda = 1$ TeV.}
\label{eventtab1} 
\end{center}
\end{table}
%
\begin{table}[!h]
\large
\begin{center}
\tabulinesep=1.2mm
\begin{tabular}{|l|c|c|c|c|c|} 
\hline 
\hline
\multicolumn{2}{|c|}{$\sqrt{s}=160$ GeV} & 
\multicolumn{3}{|c}{~~~~~~~~~~~~$N_{\rm events}$}& \\
\hline
$M_a$ (GeV)  & $\sigma_{\rm tot}$ (fb) &      Basic cuts & $\slashed{E}< 80$ GeV &  $\slashed{M}< 10$ GeV & $\sigma_{\rm cut}$ (fb)   \\
\hline
~ ~ 10 &  ~2054~  & 511509   & 476414 &  381606 & 784       \\ 
\hline 
~ ~ 80 &  ~885~   &  950028   &  942370   &  773597 & 685  \\  
\hline
~ ~150 &  ~4.62~    &  882485   &  878570   & 818390 & 3.78 \\ 
\hline
\hline
  ~  $\gamma\gamma \nu \bar{\nu}$    & ~3089~  &  186189  &  133  & 1 & 0.0031 \\ 
\hline 
\hline
\end{tabular}\\
\caption{Cross sections (before and after cuts) and event counts out of $10^6$ simulated events, for the signal   and 
the SM $\gamma\gamma \nu \bar{\nu}$ background at  $\sqrt s = 160$ GeV (versus $M_a$), applying the C1 set of basic cuts as described in the text, followed by further sequential cut optimization. We assume  $C_{a\gamma\gamma}=C_{a\gamma\bar{\gamma}}=1$
and $\Lambda = 1$ TeV.}
\label{eventtab2} 
\end{center}
\end{table}
 
\begin{table}[!h]
\large
\begin{center}
\tabulinesep=1.2mm
\begin{tabular}{|l|c|c|c|c|c|} 
\hline 
\hline
\multicolumn{2}{|c|}{$\sqrt{s}=240$ GeV} & 
\multicolumn{3}{|c}{~~~~~~~~~~~~$N_{\rm events}$}& \\
\hline
$M_a$ (GeV)  & $\sigma_{\rm tot}$ (fb) &      Basic cuts & $\slashed{E}< 120$ GeV &  $\slashed{M}< 10$ GeV & $\sigma_{\rm cut}$ (fb)   \\
\hline
~ ~ 10 &  ~2071~   & 245025   & 229988 &  163157 &  338      \\ 
\hline 
~ ~ 80 &  ~1478~   & 954942   &  941647   &  728484 & 1077  \\  
\hline
~~ 150 &  ~473~   & 947603   &  941097   & 753879 & 356  \\ 
\hline
~~ 230 &  ~2.92~  &  912354   &  907124   & 798522 & 2.33 \\ 
\hline
\hline
  ~  $\gamma\gamma \nu \bar{\nu}$    & 1615  &  175959  &  16729  & 5 & 0.0081   \\ 
\hline 
\hline
\end{tabular}\\
\caption{Cross sections (before and after cuts) and event counts out of $10^6$ simulated events, for the signal   and 
the SM $\gamma\gamma \nu \bar{\nu}$ background at  $\sqrt s = 240$ GeV (versus $M_a$), applying the C1 set of basic cuts as described in the text, followed by further sequential cut optimization. We assume  $C_{a\gamma\gamma}=C_{a\gamma\bar{\gamma}}=1$
and $\Lambda = 1$ TeV.}
\label{eventtab3} 
\end{center}
\end{table}

\begin{figure}[H]
\begin{center}
\hskip -2cm \includegraphics[width=0.49\textwidth]{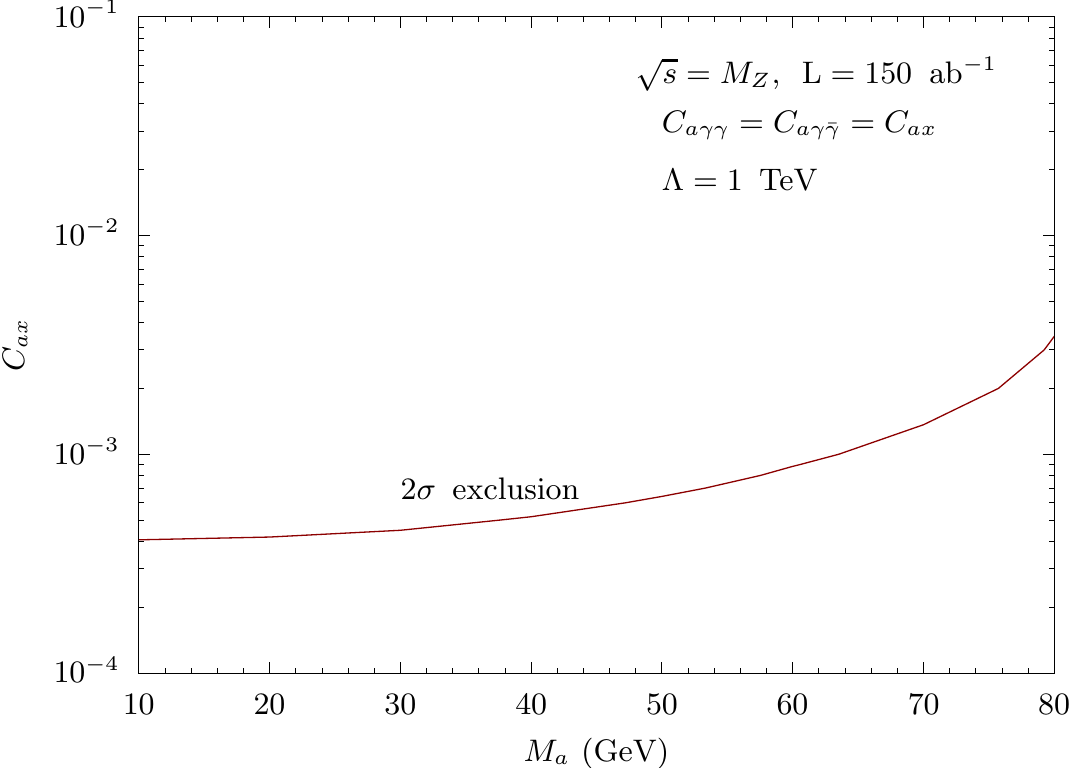} 
\caption{The $2\sigma$ exclusion limit in the  ($C_{ax},M_a$) plane 
at  $\sqrt{s}\simeq {M_{Z}}$, for an integrated {luminosity} of 150 ab$^{-1}$, assuming  $C_{a\gamma\gamma}=C_{a\gamma\bar{\gamma}}=C_{ax}$, 
and $\Lambda = 1$ TeV.   }
\label{exclplot_ecm91}
\end{center}
\vskip 0.3cm
\end{figure}
\begin{figure}[H]
\begin{center}
\hskip -2cm \includegraphics[width=0.49\textwidth]{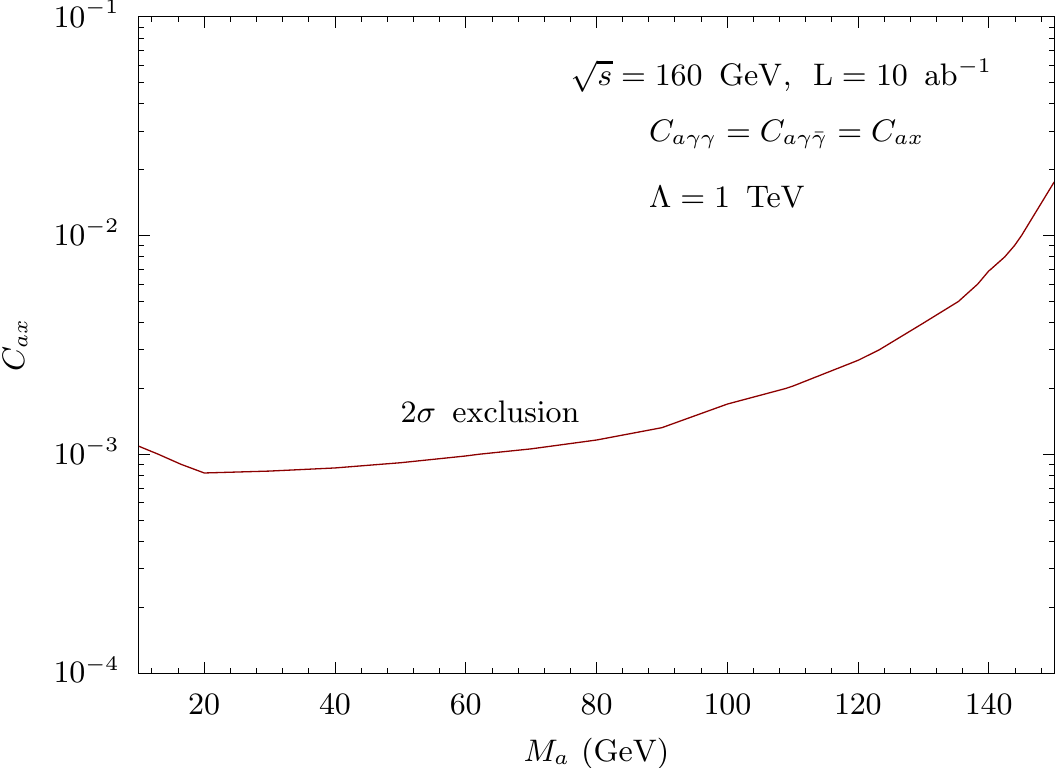} 
\caption{The $2\sigma$ exclusion limit in the  ($C_{ax},M_a$) plane  
at  $\sqrt{s}= {160}$ GeV, for an integrated {luminosity} of ${10} {\rm ~ab}^{-1}$, assuming $C_{a\gamma\gamma}=C_{a\gamma\bar{\gamma}}=C_{ax}$, 
and $\Lambda = 1$ TeV.  }
\label{exclplot_ecm160}
\end{center}
\end{figure}
\begin{figure}[H]
\begin{center}
\vskip -0.2cm
\hskip -2cm \includegraphics[width=0.49\textwidth]{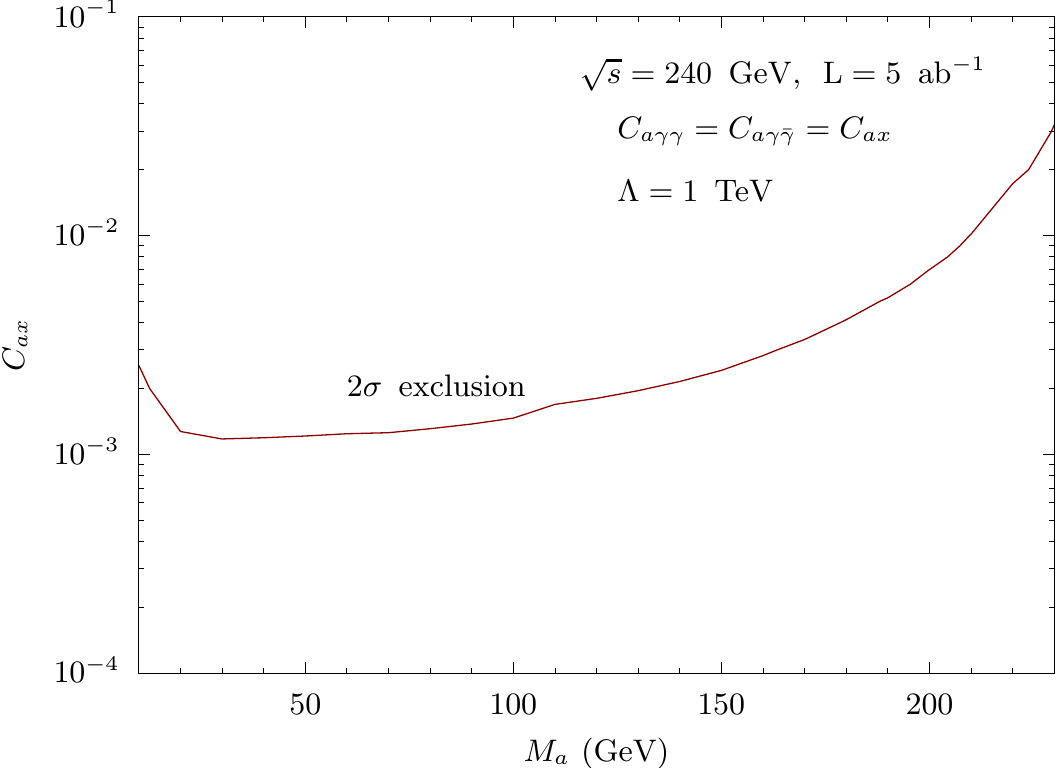} 
\caption{The $2\sigma$ exclusion limit in the  ($C_{ax},M_a$) plane 
at  $\sqrt{s}= {240}$ GeV, for an integrated {luminosity} of 5 ab$^{-1}$, assuming  $C_{a\gamma\gamma}=C_{a\gamma\bar{\gamma}}=C_{ax}$, 
and $\Lambda = 1$ TeV.  }
\label{exclplot_ecm240}
\end{center}
\end{figure}

In Figures \ref{exclplot_ecm91}, \ref{exclplot_ecm160} and \ref{exclplot_ecm240}, we show the exclusion plot in the  
$(C_{ax}, M_a)$ plane, where $C_{a\gamma\gamma}=C_{a\gamma\bar{\gamma}}=C_{ax}$, for various
 c.m. energies and corresponding expected integrated luminosities. The $2\sigma$ limit has been obtained by assuming for  
 the signal significance the following definition 
\bea
\tilde\sigma = \sqrt{2[(S+B)ln(1+S/B)-S]}
\label{eqnsignf}
\eea

where $S$ and $B$ are the numbers of observed signal and background events, respectively,
corresponding to the residual signal and background cross sections $\sigma_{\rm cut}$
in Tables \ref{eventtab1}--\ref{eventtab3}.
One can see that 
the $2\sigma$ reach for $M_a\!\!\sim$10 GeV is less sensitive compared to the one for $M_a\!\!\sim$ 20~GeV, at $\sqrt{s} =$ 160 and 240 GeV, 
despite the large
production cross section in this mass region. The decrease in sensitivity at low $M_a$ is due to 
the effect of the photon isolation requirement when the  two photons
coming from the ALP decay are mostly collimated. This effectively reduces the cut efficiency.
On the other hand, the $2\sigma$ limit on the couplings becomes less sensitive at larger 
 $M_a$ values, because of the corresponding lower production cross section.
\\

\begin{figure}[H]
\begin{center}
\hskip -2cm \includegraphics[width=0.49\textwidth]{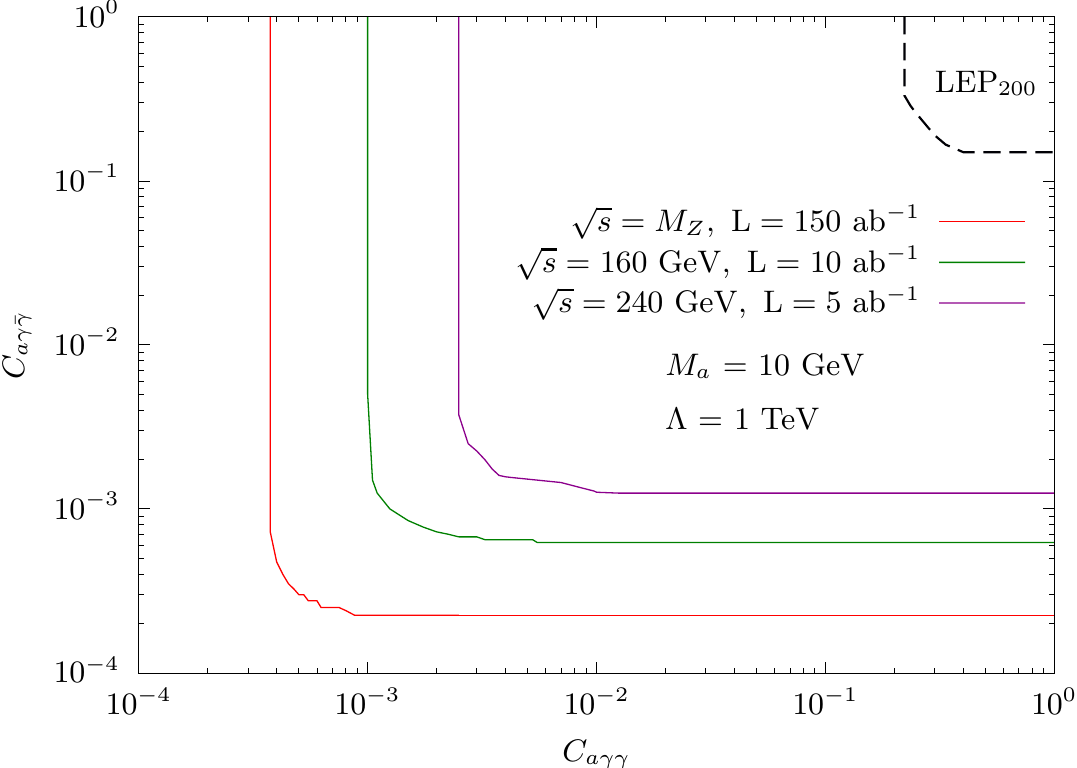} 
\hskip 0.5cm \includegraphics[width=0.49\textwidth]{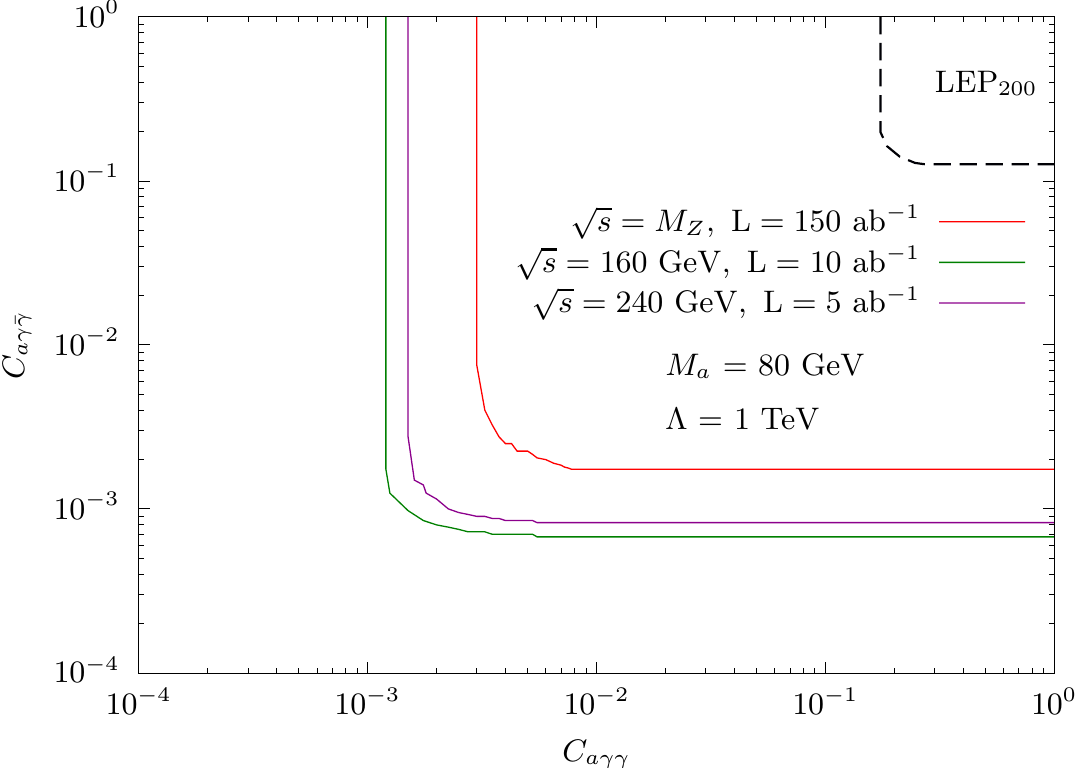} 
\caption{The $2\sigma$ exclusion limits in the  $(C_{a\gamma{\gamma}}, C_{a\gamma{\bar\gamma}})$ plane for fixed $M_a$ (left $M_a=10$ GeV, right 
$M_a=80$ GeV). The colored lines (red, green and purple) depict the bound coming from the analysis at  $\sqrt{s}\simeq {M_{Z}}$, 160~GeV,  and 240~GeV, for 
an integrated {luminosity} of 150 ab$^{-1}$, 10 ab$^{-1}$, and 5 ab$^{-1}$, respectively, and the black dashed line represents the bound coming from the 
LEP analysis as described in the text. We have assumed $\Lambda = 1$~TeV.}
\label{exclplot_10_80}
\end{center}
\end{figure}

\begin{figure}[H]
\begin{center}
\hskip -2cm \includegraphics[width=0.49\textwidth]{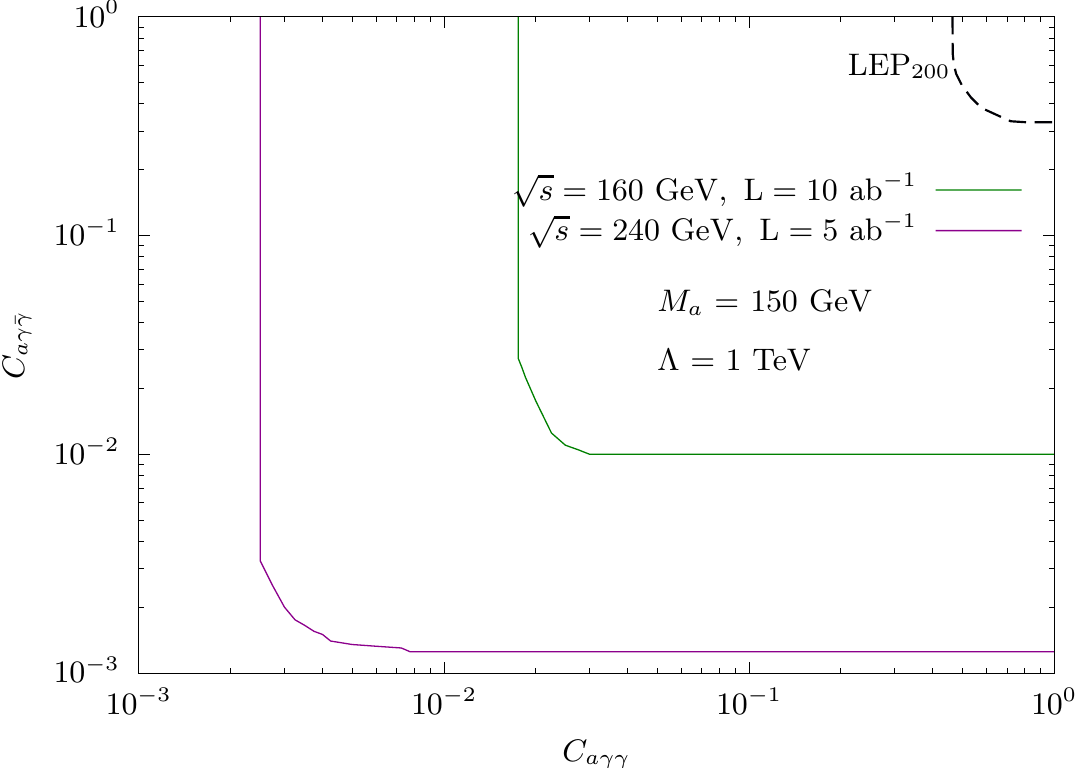} 
\hskip 0.5cm \includegraphics[width=0.49\textwidth]{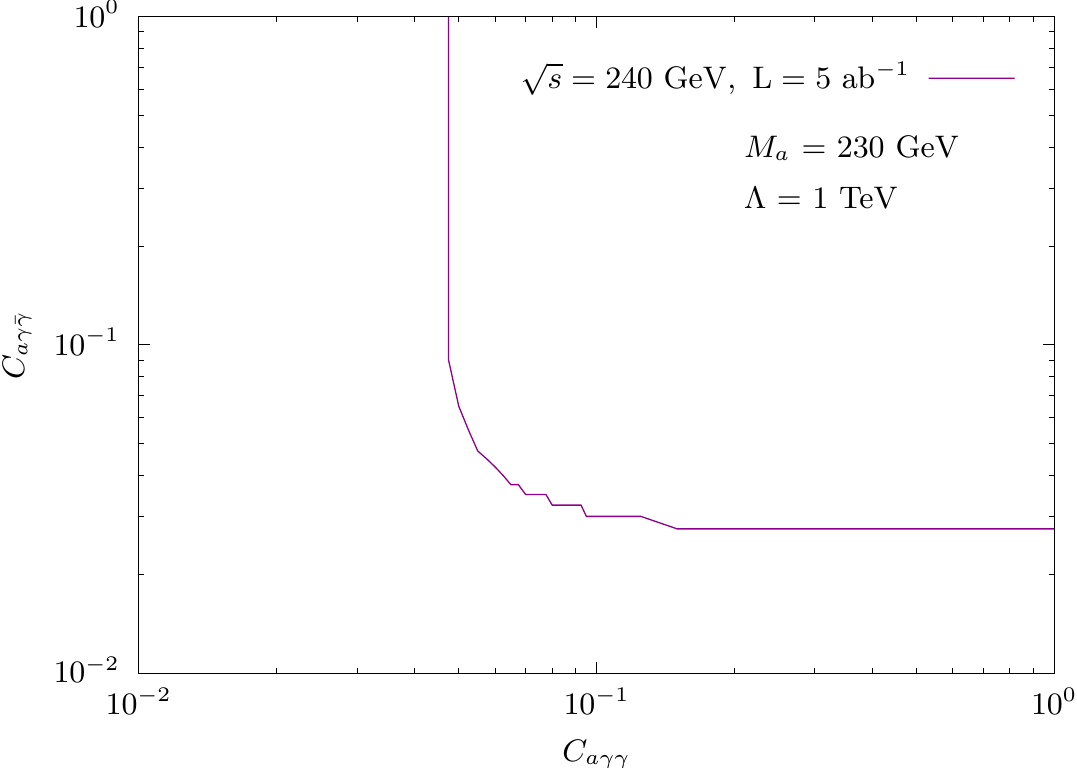} 
\caption{The $2\sigma$ exclusion limits in the  $(C_{a\gamma{\gamma}}, C_{a\gamma{\bar\gamma}})$ plane for fixed $M_a$ and given $\sqrt{s}$ (left: $M_a=150$ GeV
at  $\sqrt{s}=$ 160~GeV  and 240~GeV (green and purple lines), right:  $M_a=230$ GeV at  $\sqrt{s}=$ ~240~GeV (purple line)),  for  integrated {luminosities} as shown. 
The black dashed line in the left figure represents the bound coming from the LEP analysis 
as described in the text, for $M_a=150$ GeV. We have assumed $\Lambda = 1$~TeV.}
\label{exclplot_150_230}
\end{center}
\end{figure}

From Figures \ref{exclplot_ecm91}, \ref{exclplot_ecm160} and \ref{exclplot_ecm240}, one can see that, for masses $M_a \lsim 60$ GeV, 
the best exclusion limit is achieved at  $\sqrt{s}\simeq M_{Z}$. This is because of the very high expected integrated luminosity of about 
150  ab$^{-1}$ at this energy, despite the comparatively lower signal cross sections. 
For 60 GeV$\lsim M_a \lsim 90$ GeV, a slightly better sensitivity is obtained at 
$\sqrt{s}=160$ GeV, where a factor-two enhancement  in luminosity 
with respect to $\sqrt{s}=240$ GeV more than compensates the enhancement 
in the production cross section at higher collision energy. 
%
%
%
\vskip 0.4cm
\vskip -0.5cm
In Figures \ref{exclplot_10_80} and \ref{exclplot_150_230},
we also plot the $2\sigma$ exclusion contours in the $(C_{a\gamma{\gamma}}, C_{a\gamma\bar{\gamma}})$ plane for fixed $M_a$.  The
red, purple and green lines represent the bounds obtained from the analysis discussed in Section~\ref{secfccee}.
The black dashed curve represents the bound coming from the LEP analysis (as described in  \ref{seclep}), by assuming a null result. In these plots, 
we dubbed ${\rm LEP}_{200}$ the LEP searches in the range $\sqrt{s}= (189-208)$ GeV,
with a total integrated luminosity of 619~pb$^{-1}$~\cite{Gataullin:2005ge}. From the $2D$ plots, it is also clear that the future collider experiments will be much more sensitive, and have much better reach in the $(C_{a\gamma{\gamma}}, C_{a\gamma{\bar\gamma}})$ plane,
thanks to both a higher luminosity and to the optimized selection strategy. The significance definition in Eq.(\ref{eqnsignf}) is used everywhere to make this comparison. 
\\

\vskip 0.4cm


\vskip -0.3cm
One can notice from the plots in Figures \ref{exclplot_10_80} and \ref{exclplot_150_230}, that when one of the couplings is much lower than the other, the sensitivity to the smaller coupling is independent of the latter. 
This feature arises from the scaling of the cross section with the couplings, when ignoring interference effects. Indeed, the cross section for the
$e^+e^- \to \gamma\gamma\bar{\gamma}$ process scales as $\sigma (e^+e^- \to \gamma\gamma\bar{\gamma})  \approx C^2_{a \gamma\gamma}[\sigma_1 + \frac{\Gamma_{1}}{\Gamma_{2}}\sigma_2]$
for $C_{a \gamma\gamma}\ll C_{a \gamma\bar{\gamma}}$, while for $C_{a \gamma\bar{\gamma}}\ll C_{a \gamma\gamma}$ the scaling is given by $\sigma (e^+e^- \to \gamma\gamma\bar{\gamma})  \approx C^2_{a \gamma\bar{\gamma}}[\frac{\Gamma_{2}}{\Gamma_{1}} \sigma_1 + \sigma_2]$ [where the factors $\sigma_1  = \sigma ({e^+e^- \to \gamma a})$, $\sigma_2  = \sigma ({e^+e^- \to \bar{\gamma} a})$,
$\Gamma_1  = \Gamma ({a\to \gamma{\gamma}})$,  and $\Gamma_2  = \Gamma ({a\to \gamma\bar{\gamma}})$ are all evaluated at $C_{a\gamma\gamma} = C_{a\gamma\bar{\gamma}} = 1$ and 
$\Lambda = 1$~TeV]. Therefore the total cross section in either of these limits roughly scales with the smallest of the two couplings squared, and becomes independent of the other.

Furthermore, the sensitivity of the exclusion limit in the $(C_{a\gamma{\gamma}}, C_{a\gamma{\bar\gamma}})$ plane for fixed $M_a$ is not symmetric in the 
two couplings. The sensitivity reach  for $C_{a\gamma\bar{\gamma}}$ is indeed slightly better than that for  $C_{a\gamma\gamma}$. This is because, the partial widths $\Gamma_2$ 
and $\Gamma_1$ enter in the expression for the cross sections  describing the scaling (just defined) in inverted ratios, and $\frac{\Gamma_{2}}{\Gamma_{1}} > 1$.

Figures~\ref{exclplot_10_80} and \ref{exclplot_150_230} indeed  show that,  when there is a clear hierarchy in the 
 $a\gamma{\gamma}$ and $a\gamma\bar{\gamma}$ couplings, the $2\sigma$ exclusion dependence is restricted to  just the smallest coupling. However, in case a 
 $e^+ e^-  \to \gamma\gamma\bar{\gamma}$ signal were detected, our analysis would not 
 be able to distinguish the corresponding hierarchical  ordering of the two couplings. 

\vskip 1.0cm
\section{Summary and conclusions}
\label{conclusion}

We have considered an axion-like particle with mass in the range $(10 - 230)$ GeV, coupled to a photon and a massless dark-photon
through higher-dimensional effective operators. Such operators may occur naturally in a theory which includes  light pseudo-scalar and dark-photon particles, and  can be generated at the loop level by heavy messenger particles charged under both  $U(1)_{em}$
and $U(1)_{D}$. We have investigated the future  collider prospects for  a scenario including these effective operators. In particular, we have thoroughly 
studied  and lay down the strategy to probe a Dark-ALP portal as defined in Eq.(\ref{Leff}), at future $e^+e^-$ colliders. The conventional search for {\it three-photon} final states (which does not  include the $a\gamma\bar{\gamma}$ coupling) is replaced by the $2\gamma+\slashed{E}$ signature arising from the $e^+ e^-  \to \gamma\gamma\bar{\gamma}$ process, with the missing momentum arising from  the  dark-photon in the collision final state
well characterized by a vanishing missing mass. We assumed
 a prompt ALP decay on a typical-detector length scale. 
The main SM background coming from the  $ e^{+}e^{-} \to  \gamma\gamma \nu \bar{\nu}$ channel can then  be controlled by proper cutflows on the most sensitive kinematic variables, namely the missing energy $\slashed{E}$ and missing mass $\slashed{M}$. 
Relevant kinematical distributions and consequent selection strategies for the 
$e^+ e^-  \to \gamma\gamma\bar{\gamma}$ process have been analysed.
The missing mass variable associated to the dark photon 
turns out to be particularly efficient for the $S/B$ optimization.
Projections at future $e^+e^-$ colliders for the corresponding constraints on the $a\gamma{\gamma}$ and  
 $a\gamma\bar{\gamma}$ couplings have been discussed.
 For ALP masses not too close to the production threshold, the FCC-ee  has in general
 the potential to constrain down to $\mathcal{O} (10^{-3}-10^{-4})$ the 
ALP-photon-photon ($a\gamma{\gamma}$) and  
ALP-photon-dark-photon ($a\gamma\bar{\gamma}$) couplings, as defined in Eq.(\ref{Leff}),
 assuming  $\Lambda = 1$~TeV.
 
Note that our analysis  is not sensitive to the CP-property of the produced scalar particle, and can also be effectively used to probe  possible couplings of CP-even scalars to  photons and dark-photons, if occurring in particular new physics scenarios. 

The corresponding search strategies for the ILC, CLIC, CEPC collision-energy and integrated-luminosity setups can be inferred by the present analysis in a quite straightforward way.
\section{Acknowledgements}
E.G. wish to thank the Theoretical Physics Department at CERN for its hospitality during the completion of this work.
S.B. would like to thank the Korea Institute for Advanced Study for extending its hospitality while part of the work has been carried 
out, and also thanks E. J. Chun for useful discussions and comments. A.C. would like to thank University Grant Commission (UGC), 
India, for supporting this work by means of a NET Fellowship (Ref. No. 22/06/2014 (i) EU-V and Sr. No. 2061451168). 

\end{document}